\documentclass[11pt,twoside,a4paper,cmspaper,final,collab]{cms-tdr}
\def\svnVersion{213ca02-D}\def\svnDate{2026/08/06}\def\cmsCernNoTag{CERN-EP-2026-232}\def\cmsCernDate{\today}\def\cmsMessage{Submitted to Physical Review Letters}
\begin{document}\cmsNoteHeader{SMP-25-008}

\newcommand{\emu}{\ensuremath{{\Pepm\PGmmp}}\xspace}
\newcommand{\dr}{\ensuremath{{\Delta R}}\xspace}
\newcommand{\logroverdr}{\ensuremath{{\ln(R/\dr)}}\xspace}
\ifthenelse{\boolean{cms@external}}{\renewcommand{\appendixname}{End Matter}}{}
\providecommand{\EndMatter}{\appendixname~\ref{app:endMat}\xspace}

\cmsNoteHeader{SMP-25-008}
\title{Evidence for the dead-cone effect in bottom quark initiated jets produced in proton-proton collisions at \texorpdfstring{${\sqrt{s} = 13\TeV}$}{sqrt(s) = 13 TeV}}

\date{\today}

\abstract{
    The suppression of collinear gluon emissions from a massive quark, or the ``dead-cone effect,'' is the primary manifestation of quark mass effects in parton showers. A proton-proton collision data set recorded by the CMS experiment at $\sqrt{s}=13\TeV$ with an integrated luminosity of 59.8\fbinv is analyzed. A differential measurement is presented of the emission density of bottom quark jets, studied as a function of the angular separation between iteratively declustered subjets. The measurement is performed in bottom quark initiated jets with transverse momenta between 40 and 200\GeV originating from top quark pair production. Compared to data, a parton shower Monte Carlo model that includes the dead-cone effect is favored, while a model without the dead-cone effect is disfavored with a significance of 4.3 standard deviations.
}

\hypersetup{
pdfauthor={CMS Collaboration},
pdftitle={Direct evidence of the dead-cone effect in bottom quark initiated jets},
pdfsubject={CMS},
pdfkeywords={CMS, QCD, standard model, hadronic, dead-cone, b quark}}

\maketitle 
Jets are collimated sprays of particles produced in particle collisions at high energies.
Jet substructure studies provide stringent tests of quantum chromodynamics (QCD) and drive improvements in theoretical calculations and Monte Carlo (MC) event generators~\cite{Salam:2010nqg,Larkoski:2017jix,Kogler:2018hem,Marzani:2019hun}.
Nonzero quark masses modify the radiation pattern of jets initiated by quarks of different flavors, and require the consideration of mass-dependent effects in parton shower modeling.
Precise modeling of \PQb quark jet formation is important in many standard model processes, such as Higgs boson decays~\cite{CMS:2018nsn,CMS:2020tkr,CMS:2022cpr,CMS:2022omp,CMS:2020mpn,CMS:2024fdo,CMS:2020cga,CMS:2025dsh} and top quark production~\cite{ATLAS:2024dxp,CMS:2024irj,CMS:2021vhb,ATLAS:2022aof,CMS:2020tvq,CMS:2019esx,CMS:2023qyl}, as well as in other processes of physics beyond the standard model~\cite{ATLAS:2018tfk,CMS:2019zmd,ATLAS:2022ihe}.

\begin{figure}[htbp]
  \centering
  \ifthenelse{\boolean{cms@external}}{
    \includegraphics[width=0.48\textwidth]{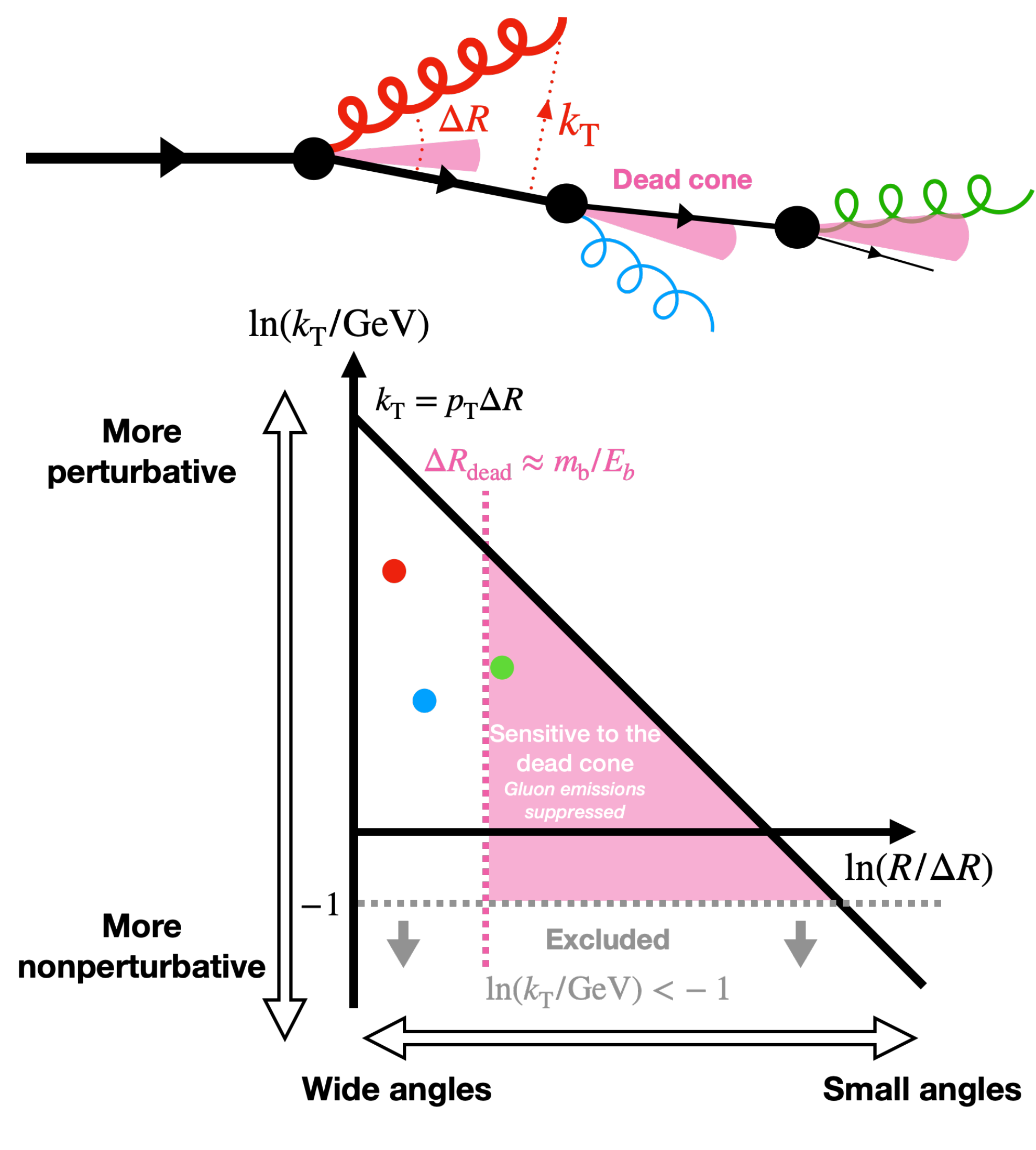}
  }{
    \includegraphics[width=0.78\textwidth]{Figure_001.pdf}
  }
	\caption{An illustration of the iterative declustering of a \PQb quark jet to construct the Lund jet plane (LJP).
		The upper diagram shows the showering of a \PQb quark (pointing arrow) ordered in descending angular separation between the \PQb quark and the gluon radiation (curly line).
		The pink cones along the direction of the emitters represent the dead cones.
		The lower diagram shows the corresponding LJP, where each colored point corresponds to a declustering step in the upper diagram and the pink block indicates the region sensitive to the dead cone.
  }
  \label{fig:ljp_illus} 
\end{figure}

In leading order perturbative QCD, the probability to have a gluon emitted nearly collinear to the radiating quark is proportional to ${\theta^2\rd\theta^2/[\theta^2+(m_{\PQq}/E_{\PQq})^2]^2}$, where $\theta$ is the angle between the quark and the gluon emission, $E_{\PQq}$ is the energy of the quark before the emission, and $m_{\PQq}$ is the mass of the quark.
Emissions within the angle of ${\Delta R_{\text{dead}}=m_{\PQq}/E_{\PQq}}$ along the emitter are suppressed~\cite{Dokshitzer:1991fd}, and this angle decreases with the emitter energy.
With the \PQb (charm) quark mass at about 4 (1)\GeV~\cite{ParticleDataGroup:2024cfk}, a sizable suppression is expected for heavy quarks~\cite{Cunqueiro:2018jbh,Ghira:2025nym}.
This is known as the ``dead-cone effect,'' illustrated in Fig.~\ref{fig:ljp_illus} (upper diagram).
Indirect evidence of the dead-cone effect has been observed~\cite{DELPHI:2000edu,CDF:2008ixu,ATLAS:2013uet,Kluth:2023umf}, and the first direct observation was reported by the ALICE Collaboration~\cite{ALICE:2021aqk} in jets containing a $\PD^{0}$ meson.
Similar direct measurements were performed by the LHCb~\cite{LHCb:2025mcq} and CMS Collaborations~\cite{CMS:2025dsz,CMS:2025afq}.

This Letter presents a direct measurement of the dead-cone effect in \PQb quark jets from top quark pair (\ttbar) production. 
The analysis uses a proton-proton ($\Pp\Pp$) collision data set of {59.8\fbinv} recorded by the CMS experiment in 2018 at the CERN LHC at $\sqrt{s}=13\TeV$~\cite{CMS-PAS-LUM-20-001}.
Similar to previous studies~\cite{ALICE:2021aqk,LHCb:2025mcq,CMS:2025dsz,CMS:2025afq}, this measurement uses the Lund jet plane (LJP)~\cite{Andersson:1988gp,Dreyer:2018nbf} to construct a proxy for the parton shower (PS) in jets.
Several new experimental techniques are introduced.
A novel strategy is employed to reconstruct charged particles from inclusive \PQb hadron decays rather than specific decay channels.
Instead of using a \PQb jet tagging algorithm that can impose biases on the substructure measurement~\cite{ATLAS:2025drf,CMS:2017wtu} of the measured jet, we select \ttbar events to provide a high purity sample of \PQb quark jets via a ``tag-and-probe'' (T\&P) approach~\cite{Khachatryan:2010xn,CMS:2018rym,CMS:2020uim,ATLAS:2022miz}.

The LJP maps different physical effects onto different regions of the plane. 
Experimental measurements of LJP distributions have been reported in Refs.~\cite{ATLAS:2020bbn,Havener:2021yhb,CMS:2023lpp,ATLAS:2024wrd,LHCb:2025mcq}.
The LJP is constructed by reclustering the constituents of anti-\kt jets~\cite{Cacciari:2008gp} in an ascending angular order with the Cambridge--Aachen (CA) algorithm~\cite{Dokshitzer:1997in}.
Then, the angle-ordered CA clustering history is traversed backwards, with each declustering step splitting a jet or a subjet into two subjets.
The subjet with the higher (lower) transverse momentum, \pt, is labeled as a proxy for a parton ``emitter'' (``emission'').
The angular separation, $\dr \equiv \sqrt{\smash[b]{(\Delta y)^2+(\Delta \phi)^2}}$, and relative transverse momentum, $\kt\equiv\pt^{\text{em}}\dr$, are measured between the emitter and emission, where $\Delta y$ ($\Delta \phi$) is the rapidity (azimuthal angle) separation, and $\pt^{\text{em}}$ is the emission-\pt.
This procedure is repeated on the emitter branch until the higher \pt subjet is composed of one constituent.
The LJP construction procedure is illustrated in Fig.~\ref{fig:ljp_illus}.
We measure the differential emission density per jet, $(1/N_{\text{jet}})\rd{N_{\text{em}}}$, along the angular ordinate \logroverdr, where $N_{\text{jet}}$ ($N_{\text{em}}$) is the number of jets measured (emissions within the measured jets) and $R=0.4$ is the distance parameter of the anti-\kt jets.
The probability of QCD radiation scales approximately as \logroverdr, in the collinear limit.
The dead-cone effect is assessed by comparing $\rho(\dr)\equiv(1/N_{\text{jet}})\rd{N_{\text{em}}}/\rd\logroverdr$ in \PQb jets to a model where the effect is absent.
A lower \kt threshold is applied to suppress nonperturbative contributions~\cite{Cunqueiro:2018jbh}.
The measured distributions are corrected for detector effects and compared with predictions from several event generators.
The results provide important benchmarks for PS development studies~\cite{Dasgupta:2020fwr} and QCD calculations~\cite{Lifson:2020gua,Ghira:2025nym}.
A \textsc{Rivet} routine is available for this measurement~\cite{Buckley:2010ar}, and the data have been made publicly available in the HEPData repository~\cite{hepdata}.

The CMS apparatus~\cite{CMS:2008xjf,CMS:2023gfb} is a multipurpose, nearly hermetic detector, designed to trigger on~\cite{CMS:2020cmk,CMS:2016ngn,CMS:2024aqx} and identify electrons, muons, photons, and (charged and neutral) hadrons~\cite{CMS:2020uim,CMS:2018rym,CMS:2014pgm}.
A global ``particle-flow'' (PF) algorithm~\cite{CMS:2017yfk} aims to reconstruct all individual particles in an event, combining information provided by the all-silicon tracker, the crystal electromagnetic calorimeter, and the brass and scintillator hadron calorimeter, operating inside a 3.8\unit{T} superconducting solenoid, with data from the gas-ionization muon detectors interleaved with the layers of the flux-return yoke outside the solenoid.
Data passing a single-muon trigger incorporating muon isolation criteria were recorded, and were required to satisfy quality criteria that reject events from beam-related backgrounds and detector noise~\cite{CMS:2016ngn}.
The primary vertex (PV) is taken to be the vertex corresponding to the hardest scattering in the event, evaluated using tracking information alone, as described in Section 9.4.1 of Ref.~\cite{CMS-TDR-15-02}.
Other vertices, resulting from additional $\Pp\Pp$ interactions occurring in the same or adjacent bunch crossings, are considered as pileup (PU) vertices.
Tracks associated with PU vertices are discarded and an offset correction is applied to correct for remaining PU contributions to the jet momentum~\cite{CMS:2020ebo}.
Secondary vertices (SV) of hadron decays are reconstructed from tracks displaced from the PV~\cite{CMS:2011yuk}.
Jets are clustered from all PF reconstructed particles using the anti-\kt algorithm~\cite{Cacciari:2008gp, Cacciari:2011ma} with a distance parameter of $R=0.4$, and calibrated using simulation-based and in situ jet energy corrections~\cite{CMS:2016lmd}.
At the generator level, jets are clustered from particles with a proper decay length longer than 10\mm.
The jet flavor is determined by the ghost association procedure~\cite{Cacciari:2007fd}; jets containing at least one ghost \PQb quark are labeled as \PQb quark jets.
A boosted decision tree selection is applied to suppress jets initiated by PU vertices~\cite{CMS:2020ebo}.
Owing to the approximate isospin symmetry of QCD~\cite{ATLAS:2024jrp}, the LJPs constructed from either charged or all hadrons are expected to be similar~\cite{Lifson:2020gua}.
Only charged jet constituents with $\abs{\eta}<2.5$ and $\pt>1\GeV$ are considered in this measurement, for both generator and detector level.
To further suppress particles from PU vertices, detector-level constituents are required to have both their transverse and longitudinal impact parameters relative to the PV less than 2\mm.

Simulated samples are used to unfold the detector effects and to compare with the measured data.
The signal \ttbar and background top quark plus \PW boson ($\PQt\PW$) processes are generated at next-to-leading order (NLO) in QCD with {\POWHEG v2.0}~\cite{Nason:2004rx, Frixione:2007vw, Alioli:2010xd, Alioli:2010xa, Alioli:2008tz, Bagnaschi:2011tu,Heinrich:2017kxx, Buchalla:2018yce}.
The Drell--Yan background is simulated at NLO with up to two additional partons using {\MGvATNLO v2.6.5} and the FxFx jet-merging procedure~\cite{Frederix:2012ps}.
The samples are interfaced with {\PYTHIA v8.240}~\cite{Sjostrand:2014zea} using the CP5 tune~\cite{Sirunyan:2019dfx} for the PS, hadronization, and underlying event modeling.
All samples use the NNPDF 3.1 next-to-NLO parton distribution functions (PDFs)~\cite{Ball:2017nwa}.
Alternative \ttbar samples are produced using the {\HERWIG v7.2.2} generator~\cite{Bellm:2015jjp} with the CH3 tune~\cite{CMS:2020dqt}.
To isolate the dead-cone effect, an additional \HERWIG\ \ttbar sample is produced with the \PQb quark mass set to zero during the PS to use a massless showering kernel, using the same CH3 tune.
The \PQb quarks are subsequently brought on-shell after the PS by redistributing the momentum of their color-connected partons while conserving the total momentum as in Section 4.3.5 of Ref.~\cite{Bellm:2025pcw}.
The kinematic information from the nominal scattering process and hadronization are preserved.
Simulated samples are processed through a \GEANTfour~\cite{Agostinelli:2002hh} simulation of the CMS detector and reconstructed in the same way as the data.
All samples include the effects of PU, and the PU multiplicity distribution in simulation is reweighted to better describe the data.

A selection of events with exactly one electron and one muon of opposite charge is applied to target \ttbar events in the \emu final state (``dileptonic \ttbar'').
The electron (muon) is required to have $\pt>20\GeV$ and $\abs{\eta}<2.5$ ($\pt>26\GeV$ and $\abs{\eta}<2.4$).
Electrons in the barrel-endcap transition region in $1.44<\abs{\eta}<1.57$ are excluded.
The events are also required to have at least two jets with $\pt>20\GeV$ and $\abs{\eta}<2.5$.
At least one of the two jets leading in \pt must be identified as a \PQb quark jet by the \textsc{DeepJet} algorithm~\cite{Bols:2020bkb}, using an efficiency (mistag rate) working point of {56\%} (0.1\%) for \PQb quark (light-quark and gluon) jets.
About {96\%} of the selected events in simulation are from dileptonic \ttbar, with the dominant background from $\PQt\PW$ production, amounting to {3.3\%}.

\begin{figure}[htbp]
  \centering
  \ifthenelse{\boolean{cms@external}}{
      \includegraphics[width=0.48\textwidth]{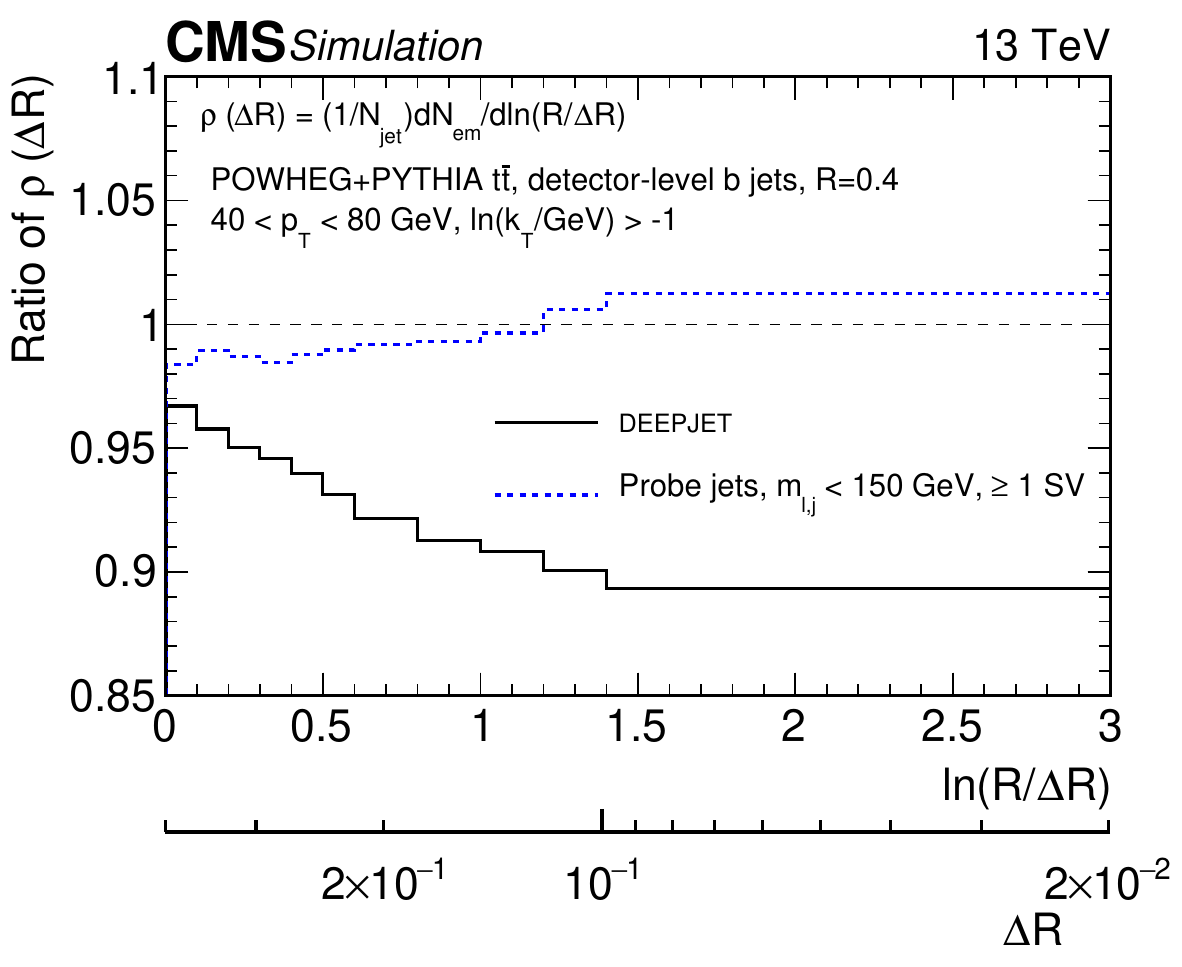}\label{fig:selection_bias}
      \includegraphics[width=0.48\textwidth]{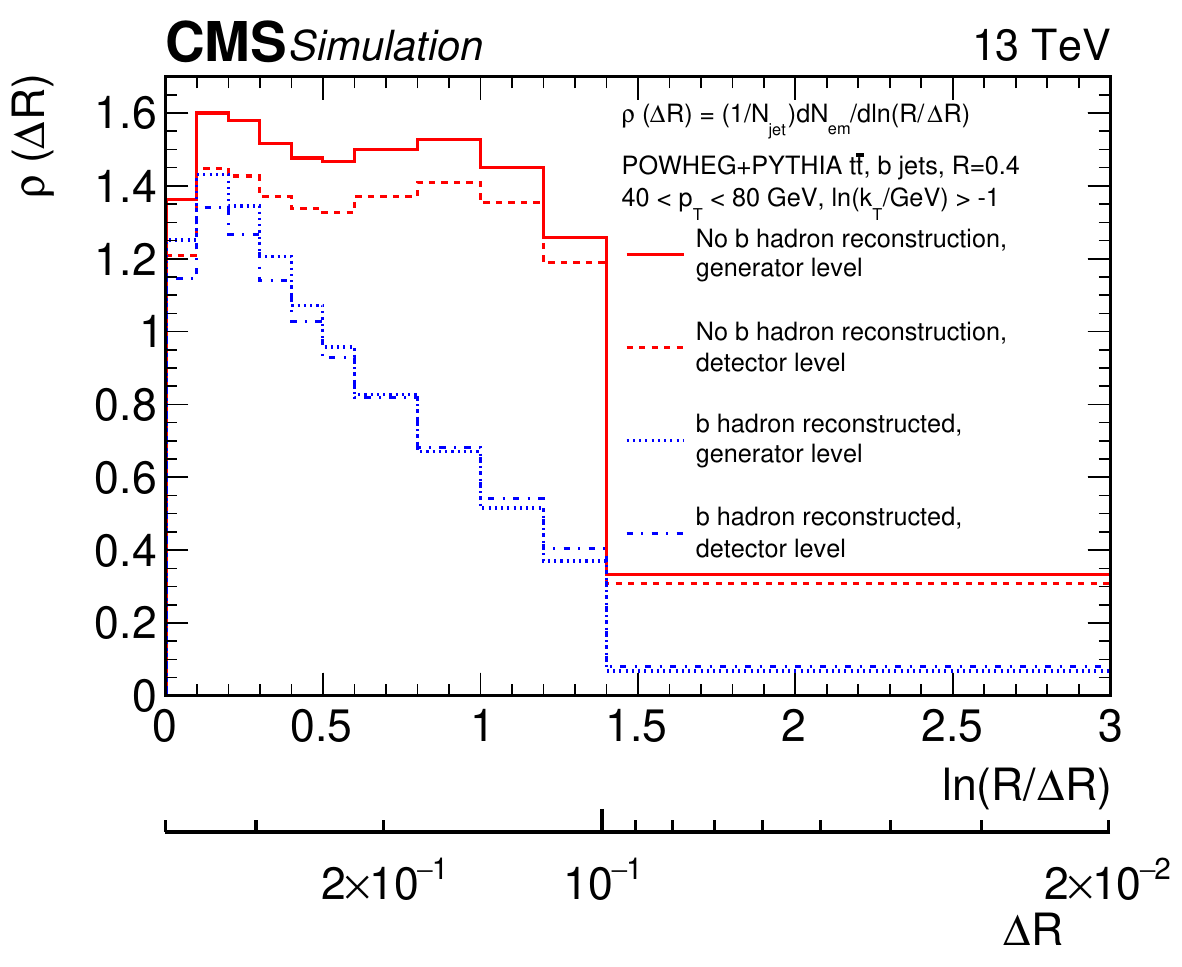}\label{fig:decay_effects}
  }{
      \includegraphics[width=0.75\textwidth]{Figure_002-a.pdf}\label{fig:selection_bias}
      \includegraphics[width=0.75\textwidth]{Figure_002-b.pdf}\label{fig:decay_effects}
  }
	\caption{
    (Upper) Ratio of the emission density distribution between the \PQb quark jets selected using either the \textsc{DeepJet} \PQb tagger or the T\&P method to the \PQb quark jets selected using generator-level information.
    (Lower) The detector-level and generator-level emission density distributions for \PQb quark jets with $40<\pt<80\GeV$ in the \POWHEG{}+\PYTHIA\ \ttbar sample are shown with and without the partial \PQb hadron reconstruction, either with generator-level information or with the transformer encoder.
    The statistical uncertainties are too small to be visible in either panel.
    } 
  \label{fig:two} 
\end{figure}

A T\&P approach selects the \PQb quark jets to be measured. 
The two jets leading in \pt are considered as a T\&P pair, where the jet passing the \textsc{DeepJet} tagger is considered as a ``tag'' and the other jet is considered as a ``probe''.
The probe is required to consist of at least two charged particles with $\abs{\eta}<2.5$ and $\pt>1\GeV$.
The probe must be within $\abs{\eta}<2$ to ensure all its constituents are within the tracker.
Only the LJP of the probe, which is not required to pass the \textsc{DeepJet} tagger, is measured.
If both leading jets pass the \textsc{DeepJet} tagger, both are measured.
To reduce light-quark and gluon contamination in the probe jet sample, additional selections are placed.
First, the probe is required to contain at least one SV.
Then, each jet of the T\&P pair is matched to one of the selected \emu by minimizing $m^2_{\ell_{i}, j_{1}}+m^2_{\ell_{j}, j_{2}}$, and the probe is required to have $m_{\ell,j}<150\GeV$.
The requirement on $m_{\ell,j}$ suppresses background jets that do not decay from a top quark, which tend to have a higher $m_{\ell,j}$ value.
In the \POWHEG{}+\PYTHIA dileptonic \ttbar simulation, about {91\%} of the selected probes originate from \PQb quarks.
Figure~\ref{fig:two}~(upper) shows the ratio of the normalized emission distribution from \PQb quark jets selected from the \textsc{DeepJet} and the T\&P methods to the \PQb jets selected using generator-level information.
The T\&P result agrees with the distribution within 2\%, while a direct selection with \textsc{DeepJet} differs by 3--10\%, with the largest deviations in large \logroverdr, the region that is most sensitive to the dead cone.

The decay products of \PQb hadrons do not obey the approximate angle ordering of QCD radiations and modify the jet representation on the LJP~\cite{Ghira:2025nym}.
To mitigate this effect, a transformer encoder (TFE) algorithm~\cite{Vaswani:2017lxt} identifies charged decay products of \PQb hadrons and sums their four-momenta to partially reconstruct the charged component of the \PQb hadron momentum, following a similar procedure in Ref.~\cite{ATLAS:2022miz,CMS:2025dsz}.
The signal identification efficiency (background rejection rate) of the model is 95.6\% (91.2\%) per charged constituent.
The effect of \PQb hadron decays on the measurement is shown in Fig.~\ref{fig:two}~(lower) by comparing the distribution with and without summing up the \PQb hadron decay products at the generator and detector levels.
Without \PQb hadron reconstruction, the decay products distort the expected smoothly falling distribution.
The distribution recovered by the TFE at the detector level follows the generator-level distribution with summation.
Further details of the TFE are provided in \EndMatter.

We employ regularized unfolding using the iterative d'Agostini method~\cite{DAgostini:1994fjx} implemented in \textsc{RooUnfold}~\cite{Brenner:2019lmf} to correct for detector resolution, biases, and inefficiencies in data.
The number of unfolding iterations was selected to ensure that the input distribution is statistically compatible with the unfolded distribution after being folded back to the detector level ($p\text{-value}>0.05$)~\cite{pvalue}.
Two iterations are applied.
The \POWHEG{}+\PYTHIA samples are used in the unfolding for the nominal result.
The measurement is performed in two jet-\pt bins of $40<\pt<80\GeV$ and $80<\pt<200\GeV$.
At both the generator and detector levels in the unfolding, additional jet-\pt bins of $20<\pt<40\GeV$ and $200<\pt<1000\GeV$ are kept to account for the migrations of jets outside of the measurement bins.
To probe a wide range of splitting angles and reduce sensitivity to hadronization effects, emissions satisfying $0<\logroverdr<3$ and $\ln(\kt/\GeV)>-1$ are included.
Emissions in each jet-\pt, $\ln(\kt)$, and \logroverdr bin are unfolded simultaneously, along with the jet multiplicity in each \pt bin.
The response matrix is constructed by geometrically matching detector- and generator-level jets (emissions) within $\dr<0.2\,(0.1)$.
The binning in \logroverdr is optimized to maximize the granularity of the measurement consistent with the expected detector resolution.
In this binning scheme, the fraction of detector-level emissions that are in the same bin as their generator-level match is above {50\%}, for all bins.
A bin-by-bin signal purity correction is applied before the unfolding to remove the expected fraction of jets or emissions from background processes, and detector-level jets or emissions without a generator match.
The fraction of jets before declustering (emissions on the LJP) matched to the generator level is around {84--88\%} ({56--80\%}).
Similarly, a bin-by-bin efficiency correction is applied after the unfolding to correct for the fraction of jets or emissions at the generator level that do not have a detector-level match.
The efficiency of jets (emissions) selected is around 18--20\% (11--20\%).

\begin{figure}[htbp!]
  \centering
  \ifthenelse{\boolean{cms@external}}{
    \includegraphics[width=0.48\textwidth]{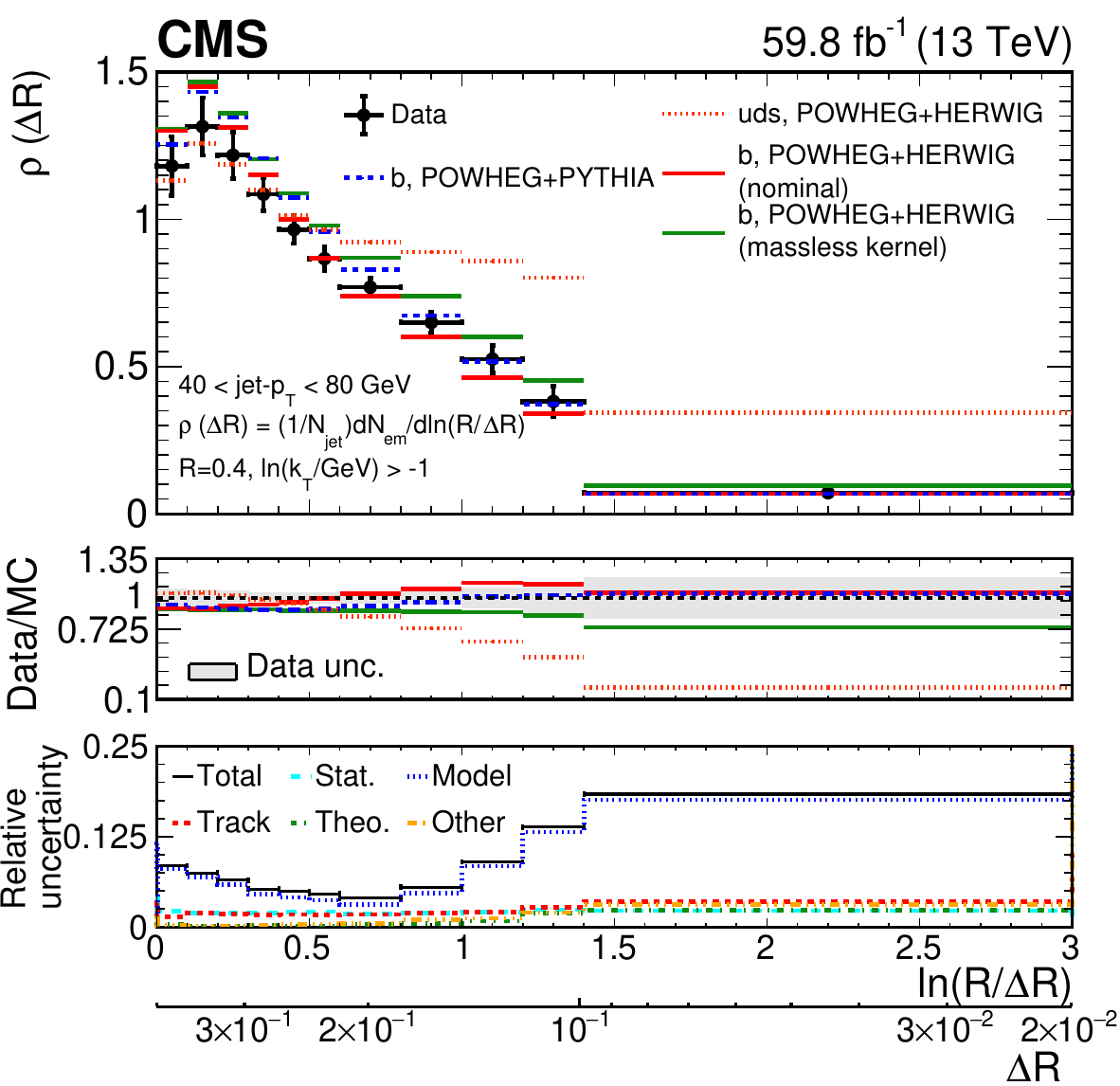}
    \includegraphics[width=0.48\textwidth]{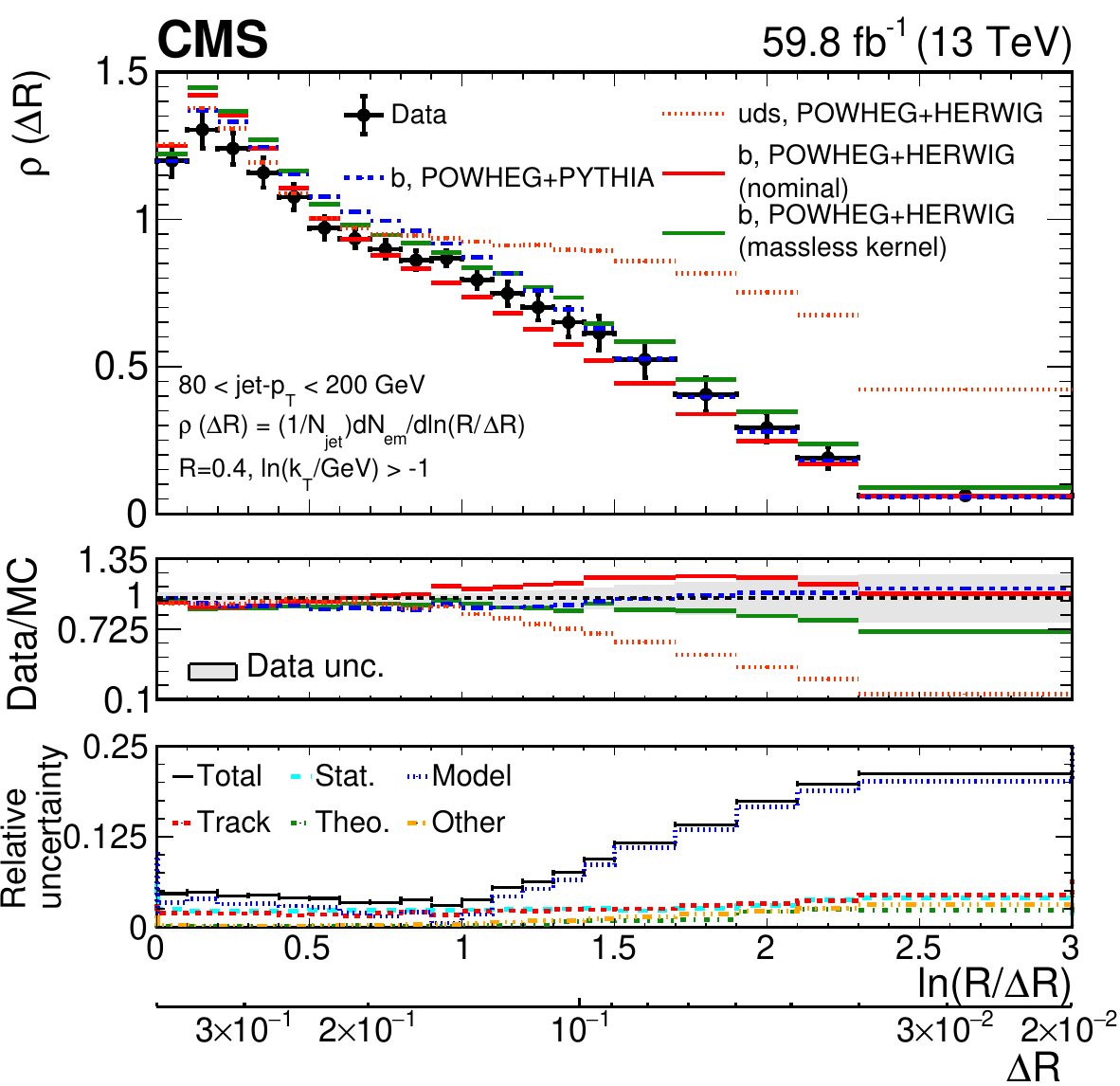}
  }{
    \includegraphics[width=0.65\textwidth]{Figure_003-a.pdf}
    \includegraphics[width=0.65\textwidth]{Figure_003-b.pdf}
  }
  \caption{The measured emission density per \logroverdr for \PQb quark jets with (upper) $40<\pt<80\GeV$ and (lower) $80<\pt<200\GeV$, with vertical error bars representing the total uncertainties.
          Data are compared with the generator-level distributions from different MC samples. 
          The middle panels show the ratio of the data to the MC predictions with a dashed line at one for reference.
          The total uncertainties of the data are shown as the gray band.
          The lower panels show the breakdown of the data uncertainties into various sources.}
  \label{fig:results} 
\end{figure}

Systematic uncertainties are evaluated by varying the nominal unfolding setup and taking the difference from the nominal result.
Physics modeling uncertainties dominate, and are assessed by replacing the nominal \POWHEG{}+\PYTHIA sample in the unfolding with \POWHEG{}+\HERWIG in three uncorrelated components: the generator-level distribution used for regularization, purity corrections, and the response matrix together with the efficiency corrections.
The generator-level and response matrix components contribute an uncertainty of up to 10\%, whereas purity corrections contribute an uncertainty in the range of 5--10\% at wide angles, increasing to 20\% in the most collinear bin.
The second largest uncertainty arises from the track reconstruction efficiency, which is higher in simulation than in data.
The discrepancy is modeled by randomly dropping 3\% of charged jet constituents in simulation~\cite{CMS-DP-2022-012}, yielding an uncertainty of 1.5--4.5\% in the measurement.
Uncertainties in initial and final-state radiation scales, factorization and renormalization scales~\cite{Butterworth:2015oua}, the PDFs and the strong coupling constant (\alpS)~\cite{PDF4LHCWorkingGroup:2022cjn}, and the top quark \pt reweighting~\cite{CMS:2016uiq,CMS:2016oae} are each below 4.5\%.
Experimental uncertainties from the jet energy scale and resolution~\cite{CMS:2016lmd}, {\PQb} jet tagging efficiency~\cite{CMS:2017wtu, CMS-DP-2018-033}, PU jet identification~\cite{CMS:2020ebo}, lepton reconstruction, identification, and trigger efficiencies~\cite{Khachatryan:2010xn}, PU vertex multiplicities~\cite{CMS:2018mlc}, and level-1 muon trigger inefficiency~\cite{CMS:2020cmk} are each below 2\%.
Uncertainties in the background cross sections and the top quark mass have a negligible impact for this measurement.
Correlations between emissions from the same jet are considered in the statistical uncertainties.
The lower panels of Fig.~\ref{fig:results} show the breakdown of uncertainties in each measurement bin.

The unfolded distributions as a function of \logroverdr for $40 < \pt < 80\GeV$ and $80 < \pt < 200\GeV$ are compared with the generator-level distributions from \PQb quark jets selected in the \POWHEG{}+\PYTHIA and \POWHEG{}+\HERWIG dileptonic \ttbar samples in Fig.~\ref{fig:results}.
For the lower jet-\pt range, deviations of around {10\%} are observed between the data and the predictions in $0<\logroverdr<0.4$.
For $0.4<\logroverdr<1.2$, \POWHEG{}+\PYTHIA improves its description of data towards large \logroverdr, while \POWHEG{}+\HERWIG begins to underestimate the number of emissions.
Both predictions agree with the data in the bin with the largest \logroverdr, or the most collinear bin.
The higher jet-\pt range shows a similar trend: the \POWHEG{}+\PYTHIA prediction shows an overestimation of emissions for $\logroverdr<1.2$, but improves towards the collinear region; the \POWHEG{}+\HERWIG prediction increasingly underestimates emissions in the intermediate region up to $\logroverdr=1.7$, but is consistent with data in the most collinear bin.

The data are also compared with the generator-level distributions from a sample of light-quark jets obtained from hadronic \PW boson decays in the \POWHEG{}+\HERWIG $\PQt\PW$ sample.
The light-quark jet sample is selected by requiring exactly one muon from the top quark decay, one \PQb quark jet with $\abs{\eta}<2.5$, and two additional light-quark jets with $\abs{\eta}<2$ from a hadronically decaying \PW boson.
The \pt spectrum of the light-quark jets sample is reweighted to match that of the selected \PQb quark jets.
In Fig.~\ref{fig:results}, the data exhibit a strong suppression of emissions relative to light-quark jets, consistent with previous observations~\cite{ALICE:2021aqk,LHCb:2025mcq,CMS:2025dsz,CMS:2025afq}. 

The use of the massless kernel sample separates the dead-cone effect from other suppression effects that may arise in heavy quark jet formulations.
The massless kernel shows higher emission densities than the nominal \PQb jet distributions in the collinear region, but to a lesser extent than the light-quark jet expectations, suggesting that hadronization effects for heavy-flavor jets play an important role.
To quantify the observed significance of the dead-cone effect, the data are compared with the nominal and the massless kernel sample generated with \POWHEG{}+\HERWIG.
The $\chi^2$ between the data and each sample is computed, and the $p$-value of the $\chi^2$ is estimated by bootstrapping~\cite{Efron:1979bxm} the MC sample at the generator level.
Both $\chi^2$ values are evaluated for $\logroverdr>0.4$ to exclude the wide angle region, which is insensitive to the dead cone effect.
For jets with $40<\pt<80\GeV$, a $p$-value of $8\times10^{-6}$, or a significance of 4.3 standard deviations, is observed relative to the massless prediction, while the nominal \POWHEG{}+\HERWIG prediction is in better agreement with data ($p$-value of 0.57).
For $80<\pt<200\GeV$, neither model shows a significant deviation from data; the nominal (massless) \POWHEG{}+\HERWIG prediction yields a $p$-value of 0.03 (0.01): sensitivity to the dead-cone effect is limited in this higher \pt range, where the dead cone is expected to be narrower.

In summary, a measurement of the emission density distribution in bottom quark initiated jets (\PQb jets) from top quark pair events in proton-proton collisions at $\sqrt{s}=13\TeV$ with the CMS detector has been presented.
A sample of \PQb jets with minimized flavor tagging biases is selected with a tag-and-probe method.
A transformer encoder is trained to partially reconstruct \PQb hadrons in jets and preserve an uncontaminated angle-ordered emission history, reconstructed on the Lund jet plane~\cite{Andersson:1988gp,Dreyer:2018nbf}.
A new experimental benchmark is provided to compare with analytical quantum chromodynamics predictions including the effect of quark masses~\cite{Lifson:2020gua,Ghira:2025nym} and for parton shower model development~\cite{Dasgupta:2020fwr}.
A suppression of collinear emissions relative to a light-quark simulation is shown by the measured emission density distribution.
Compared to data, a parton shower Monte Carlo model that includes the dead-cone effect is favored, while a model without the dead-cone effect is disfavored with a significance of 4.3 standard deviations.

\begin{acknowledgments}
We congratulate our colleagues in the CERN accelerator departments for the excellent performance of the LHC and thank the technical and administrative staffs at CERN and at other CMS institutes for their contributions to the success of the CMS effort. In addition, we gratefully acknowledge the computing centers and personnel of the Worldwide LHC Computing Grid and other centers for delivering so effectively the computing infrastructure essential to our analyses. Finally, we acknowledge the enduring support for the construction and operation of the LHC, the CMS detector, and the supporting computing infrastructure provided by the following funding agencies: SC (Armenia), BMFWF and FWF (Austria); FNRS and FWO (Belgium); CNPq, CAPES, FAPERJ, FAPERGS, and FAPESP (Brazil); MES and BNSF (Bulgaria); CERN; CAS, MoST, and NSFC (China); MINCIENCIAS (Colombia); MSES and CSF (Croatia); RIF (Cyprus); SENESCYT (Ecuador); ERC PRG and PSG, TARISTU24-TK10 and MoER TK202 (Estonia); Academy of Finland, MEC, and HIP (Finland); CEA and CNRS/IN2P3 (France); SRNSF (Georgia); BMFTR, DFG, and HGF (Germany); GSRI (Greece); MATE and NKFIH (Hungary); DAE and DST (India); IPM (Iran); SFI (Ireland); INFN (Italy); MSIT and NRF (Republic of Korea); MES (Latvia); LMTLT (Lithuania); MOE and UM (Malaysia); BUAP, CINVESTAV, CONACYT, LNS, SEP, and UASLP-FAI (Mexico); MOS (Montenegro); MBIE (New Zealand); PAEC (Pakistan); MSHE, NSC, and NAWA (Poland); FCT (Portugal); NASU (Ukraine); STFC (United Kingdom); DOE and NSF (USA).
\end{acknowledgments}\section*{Data availability} Release and preservation of data used by the CMS Collaboration as the basis for publications is guided by the  \href{https://doi.org/10.7483/OPENDATA.CMS.1BNU.8V1W}{CMS data preservation, re-use and open access policy}.

\bibliography{auto_generated} 

@ARTICLE{Salam:2010nqg,
	AUTHOR=	"Salam, Gavin P.",
	TITLE=	"Towards Jetography",
	EPRINT=	"0906.1833",
	ARCHIVEPREFIX=	"arXiv",
	PRIMARYCLASS=	"hep-ph",
	DOI=	"10.1140/epjc/s10052-010-1314-6",
	JOURNAL=	"Eur. Phys. J. C",
	VOLUME=	"67",
	PAGES=	"637",
	YEAR=	"2010",
}

@ARTICLE{Larkoski:2017jix,
	AUTHOR=	"Larkoski, Andrew J. and Moult, Ian and Nachman, Benjamin",
	TITLE=	"Jet Substructure at the {Large Hadron Collider}: a review of recent advances in theory and machine learning",
	EPRINT=	"1709.04464",
	ARCHIVEPREFIX=	"arXiv",
	PRIMARYCLASS=	"hep-ph",
	DOI=	"10.1016/j.physrep.2019.11.001",
	JOURNAL=	"Phys. Rept.",
	VOLUME=	"841",
	PAGES=	"1",
	YEAR=	"2020",
}

@ARTICLE{Kogler:2018hem,
	AUTHOR=	"Kogler, Roman and others",
	TITLE=	"{Jet substructure at the Large Hadron Collider: experimental review}",
	EPRINT=	"1803.06991",
	ARCHIVEPREFIX=	"arXiv",
	PRIMARYCLASS=	"hep-ex",
	REPORTNUMBER=	"FERMILAB-PUB-18-123-PPD",
	DOI=	"10.1103/RevModPhys.91.045003",
	JOURNAL=	"Rev. Mod. Phys.",
	VOLUME=	"91",
	PAGES=	"045003",
	YEAR=	"2019",
}

@BOOK{Marzani:2019hun,
	AUTHOR=	"Marzani, Simone and Soyez, Gregory and Spannowsky, Michael",
	TITLE=	"Looking inside jets: an introduction to jet substructure and boosted-object phenomenology",
	EPRINT=	"1901.10342",
	ARCHIVEPREFIX=	"arXiv",
	PRIMARYCLASS=	"hep-ph",
	DOI=	"10.1007/978-3-030-15709-8",
	PUBLISHER=	"Springer",
	VOLUME=	"958",
	YEAR=	"2019",
	NOTE=	"arXiv:1901.10342",
}

@ARTICLE{CMS:2018nsn,
	AUTHOR=	"{CMS Collaboration}",
	TITLE=	"{Observation of Higgs boson decay to bottom quarks}",
	EPRINT=	"1808.08242",
	ARCHIVEPREFIX=	"arXiv",
	PRIMARYCLASS=	"hep-ex",
	REPORTNUMBER=	"CMS-HIG-18-016, CERN-EP-2018-223",
	DOI=	"10.1103/PhysRevLett.121.121801",
	JOURNAL=	"Phys. Rev. Lett.",
	VOLUME=	"121",
	PAGES=	"121801",
	YEAR=	"2018",
}

@ARTICLE{CMS:2020tkr,
	AUTHOR=	"{CMS Collaboration}",
	TITLE=	"{Search for nonresonant Higgs boson pair production in final states with two bottom quarks and two photons in proton-proton collisions at $ \sqrt{s} $ = 13 TeV}",
	EPRINT=	"2011.12373",
	ARCHIVEPREFIX=	"arXiv",
	PRIMARYCLASS=	"hep-ex",
	REPORTNUMBER=	"CMS-HIG-19-018, CERN-EP-2020-222",
	DOI=	"10.1007/JHEP03(2021)257",
	JOURNAL=	"JHEP",
	VOLUME=	"03",
	PAGES=	"257",
	YEAR=	"2021",
}

@ARTICLE{CMS:2022cpr,
	AUTHOR=	"{CMS Collaboration}",
	TITLE=	"{Search for Higgs boson pair production in the four b quark final state in proton-proton collisions at $ \sqrt{s} $ = 13 TeV}",
	EPRINT=	"2202.09617",
	ARCHIVEPREFIX=	"arXiv",
	PRIMARYCLASS=	"hep-ex",
	REPORTNUMBER=	"CMS-HIG-20-005, CERN-EP-2022-004",
	DOI=	"10.1103/PhysRevLett.129.081802",
	JOURNAL=	"Phys. Rev. Lett.",
	VOLUME=	"129",
	PAGES=	"081802",
	YEAR=	"2022",
}

@ARTICLE{CMS:2022omp,
	AUTHOR=	"{CMS Collaboration}",
	TITLE=	"{Search for nonresonant Higgs boson pair production in the four leptons plus two b jets final state in proton-proton collisions at $ \sqrt{s} $ = 13 TeV}",
	EPRINT=	"2206.10657",
	ARCHIVEPREFIX=	"arXiv",
	PRIMARYCLASS=	"hep-ex",
	REPORTNUMBER=	"CMS-HIG-20-004, CERN-EP-2022-114",
	DOI=	"10.1007/JHEP06(2023)130",
	JOURNAL=	"JHEP",
	VOLUME=	"06",
	PAGES=	"130",
	YEAR=	"2023",
}

@ARTICLE{CMS:2020mpn,
	AUTHOR=	"{CMS Collaboration}",
	TITLE=	"{Measurement of the Higgs boson production rate in association with top quarks in final states with electrons, muons, and hadronically decaying tau leptons at $\sqrt{s} =$ 13 TeV}",
	EPRINT=	"2011.03652",
	ARCHIVEPREFIX=	"arXiv",
	PRIMARYCLASS=	"hep-ex",
	REPORTNUMBER=	"CMS-HIG-19-008, CERN-EP-2020-200",
	DOI=	"10.1140/epjc/s10052-021-09014-x",
	JOURNAL=	"Eur. Phys. J. C",
	VOLUME=	"81",
	PAGES=	"378",
	YEAR=	"2021",
}

@ARTICLE{CMS:2024fdo,
	AUTHOR=	"{CMS Collaboration}",
	TITLE=	"{Measurement of the $ \textrm{t}\overline{\textrm{t}}\textrm{H} $ and tH production rates in the H $\rightarrow \textrm{b}\overline{\textrm{b}} $ decay channel using proton-proton collision data at $ \sqrt{s} $ = 13 TeV}",
	EPRINT=	"2407.10896",
	ARCHIVEPREFIX=	"arXiv",
	PRIMARYCLASS=	"hep-ex",
	REPORTNUMBER=	"CMS-HIG-19-011, CERN-EP-2024-179",
	DOI=	"10.1007/JHEP02(2025)097",
	JOURNAL=	"JHEP",
	VOLUME=	"02",
	PAGES=	"097",
	YEAR=	"2025",
}

@ARTICLE{CMS:2020cga,
	AUTHOR=	"{CMS Collaboration}",
	TITLE=	"{Measurements of $\mathrm{t\bar{t}}\mathrm{H}$ production and the CP structure of the Yukawa interaction between the Higgs boson and top quark in the diphoton decay channel}",
	EPRINT=	"2003.10866",
	ARCHIVEPREFIX=	"arXiv",
	PRIMARYCLASS=	"hep-ex",
	REPORTNUMBER=	"CMS-HIG-19-013, CERN-EP-2020-028",
	DOI=	"10.1103/PhysRevLett.125.061801",
	JOURNAL=	"Phys. Rev. Lett.",
	VOLUME=	"125",
	PAGES=	"061801",
	YEAR=	"2020",
}

@ARTICLE{CMS:2025dsh,
	AUTHOR=	"{CMS Collaboration}",
	TITLE=	"{Simultaneous probe of the charm and bottom quark Yukawa couplings using ttH events}",
	EPRINT=	"2509.22535",
	ARCHIVEPREFIX=	"arXiv",
	PRIMARYCLASS=	"hep-ex",
	REPORTNUMBER=	"CMS-HIG-24-018, CERN-EP-2025-202",
	DOI=	"10.1103/9nwb-splk",
	JOURNAL=	"Phys. Rev. Lett.",
	VOLUME=	"136",
	PAGES=	"011801",
	YEAR=	"2026",
}

@ARTICLE{ATLAS:2024dxp,
	AUTHOR=	"Hayrapetyan, Aram and others",
	COLLABORATION=	"ATLAS, CMS",
	TITLE=	"{Combination of measurements of the top quark mass from data collected by the ATLAS and CMS experiments at $ \sqrt{s} $ = 7 and 8 TeV}",
	EPRINT=	"2402.08713",
	ARCHIVEPREFIX=	"arXiv",
	PRIMARYCLASS=	"hep-ex",
	REPORTNUMBER=	"CMS-TOP-22-001, ATLAS-TOPQ-2019-13, CERN-EP-2024-020",
	DOI=	"10.1103/PhysRevLett.132.261902",
	JOURNAL=	"Phys. Rev. Lett.",
	VOLUME=	"132",
	PAGES=	"261902",
	YEAR=	"2024",
}

@ARTICLE{CMS:2024irj,
	AUTHOR=	"{CMS Collaboration}",
	TITLE=	"{Review of top quark mass measurements in CMS}",
	EPRINT=	"2403.01313",
	ARCHIVEPREFIX=	"arXiv",
	PRIMARYCLASS=	"hep-ex",
	REPORTNUMBER=	"CMS-TOP-23-003, CERN-EP-2024-005",
	DOI=	"10.1016/j.physrep.2024.12.002",
	JOURNAL=	"Phys. Rept.",
	VOLUME=	"1115",
	PAGES=	"116",
	YEAR=	"2025",
}

@ARTICLE{CMS:2021vhb,
	AUTHOR=	"{CMS Collaboration}",
	TITLE=	"{Measurement of differential $\mathrm{t \bar t}$ production cross sections in the full kinematic range using lepton+jets events from proton-proton collisions at $\sqrt {s}$ = 13{\,}{\,}TeV}",
	EPRINT=	"2108.02803",
	ARCHIVEPREFIX=	"arXiv",
	PRIMARYCLASS=	"hep-ex",
	REPORTNUMBER=	"CMS-TOP-20-001, CERN-EP-2021-135",
	DOI=	"10.1103/PhysRevD.104.092013",
	JOURNAL=	"Phys. Rev. D",
	VOLUME=	"104",
	PAGES=	"092013",
	YEAR=	"2021",
}

@ARTICLE{ATLAS:2022aof,
	AUTHOR=	"Aad, G. and others",
	COLLABORATION=	"ATLAS, CMS",
	TITLE=	"{Combination of inclusive top quark pair production cross section measurements using ATLAS and CMS data at $ \sqrt{s} $ = 7 and 8 TeV}",
	EPRINT=	"2205.13830",
	ARCHIVEPREFIX=	"arXiv",
	PRIMARYCLASS=	"hep-ex",
	REPORTNUMBER=	"CERN-EP-2021-222, CMS-TOP-18-014, CERN-LPCC-2022-03",
	DOI=	"10.1007/JHEP07(2023)213",
	JOURNAL=	"JHEP",
	VOLUME=	"07",
	PAGES=	"213",
	YEAR=	"2023",
}

@ARTICLE{CMS:2020tvq,
	AUTHOR=	"{CMS Collaboration}",
	TITLE=	"{Measurement of differential $\mathrm{t\bar{t}}$ production cross sections using top quarks at large transverse momenta in pp collisions at $\sqrt{s} =$ 13 TeV}",
	EPRINT=	"2008.07860",
	ARCHIVEPREFIX=	"arXiv",
	PRIMARYCLASS=	"hep-ex",
	REPORTNUMBER=	"CMS-TOP-18-013, CERN-EP-2020-121",
	DOI=	"10.1103/PhysRevD.103.052008",
	JOURNAL=	"Phys. Rev. D",
	VOLUME=	"103",
	PAGES=	"052008",
	YEAR=	"2021",
}

@ARTICLE{CMS:2019esx,
	AUTHOR=	"{CMS Collaboration}",
	TITLE=	"{Measurement of $\mathrm{t\bar t}$ normalized multidifferential cross sections in pp collisions at $\sqrt s=13$ TeV, and simultaneous determination of the strong coupling strength, top quark pole mass, and parton distribution functions}",
	EPRINT=	"1904.05237",
	ARCHIVEPREFIX=	"arXiv",
	PRIMARYCLASS=	"hep-ex",
	REPORTNUMBER=	"CMS-TOP-18-004, CERN-EP-2019-028",
	DOI=	"10.1140/epjc/s10052-020-7917-7",
	JOURNAL=	"Eur. Phys. J. C",
	VOLUME=	"80",
	PAGES=	"658",
	YEAR=	"2020",
}

@ARTICLE{CMS:2023qyl,
	AUTHOR=	"{CMS Collaboration}",
	TITLE=	"First measurement of the top quark pair production cross section in proton-proton collisions at $ \sqrt{s} $ = 13.6 {TeV}",
	EPRINT=	"2303.10680",
	ARCHIVEPREFIX=	"arXiv",
	PRIMARYCLASS=	"hep-ex",
	REPORTNUMBER=	"CMS-TOP-22-012, CERN-EP-2023-021",
	DOI=	"10.1007/JHEP08(2023)204",
	JOURNAL=	"JHEP",
	VOLUME=	"08",
	PAGES=	"204",
	YEAR=	"2023",
}

@ARTICLE{ATLAS:2018tfk,
	AUTHOR=	"{ATLAS Collaboration}",
	TITLE=	"{Search for resonances in the mass distribution of jet pairs with one or two jets identified as b jets in proton-proton collisions at $\sqrt{s}=13$ TeV with the ATLAS detector}",
	EPRINT=	"1805.09299",
	ARCHIVEPREFIX=	"arXiv",
	PRIMARYCLASS=	"hep-ex",
	REPORTNUMBER=	"CERN-EP-2018-075",
	DOI=	"10.1103/PhysRevD.98.032016",
	JOURNAL=	"Phys. Rev. D",
	VOLUME=	"98",
	PAGES=	"032016",
	YEAR=	"2018",
}

@ARTICLE{CMS:2019zmd,
	AUTHOR=	"{CMS Collaboration}",
	TITLE=	"{Search for supersymmetry in proton-proton collisions at 13 TeV in final states with jets and missing transverse momentum}",
	EPRINT=	"1908.04722",
	ARCHIVEPREFIX=	"arXiv",
	PRIMARYCLASS=	"hep-ex",
	REPORTNUMBER=	"CMS-SUS-19-006, CERN-EP-2019-152",
	DOI=	"10.1007/JHEP10(2019)244",
	JOURNAL=	"JHEP",
	VOLUME=	"10",
	PAGES=	"244",
	YEAR=	"2019",
}

@ARTICLE{ATLAS:2022ihe,
	AUTHOR=	"{ATLAS Collaboration}",
	TITLE=	"{Search for supersymmetry in final states with missing transverse momentum and three or more b jets in 139 fb$^{-1}$ of proton{\textendash}proton collisions at $\sqrt{s} = 13$~TeV with the ATLAS detector}",
	EPRINT=	"2211.08028",
	ARCHIVEPREFIX=	"arXiv",
	PRIMARYCLASS=	"hep-ex",
	REPORTNUMBER=	"CERN-EP-2022-213",
	DOI=	"10.1140/epjc/s10052-023-11543-6",
	JOURNAL=	"Eur. Phys. J. C",
	VOLUME=	"83",
	PAGES=	"561",
	YEAR=	"2023",
}

@ARTICLE{Dokshitzer:1991fd,
	AUTHOR=	"Dokshitzer, Yuri L. and Khoze, Valery A. and Troian, S. I.",
	TITLE=	"{On specific {QCD} properties of heavy quark fragmentation ('dead cone')}",
	DOI=	"10.1088/0954-3899/17/10/023",
	JOURNAL=	"J. Phys. G",
	VOLUME=	"17",
	PAGES=	"1602",
	YEAR=	"1991",
}

@ARTICLE{ParticleDataGroup:2024cfk,
	AUTHOR=	"Navas, S. and others",
	COLLABORATION=	"Particle Data Group",
	TITLE=	"Review of particle physics",
	DOI=	"10.1103/PhysRevD.110.030001",
	JOURNAL=	"Phys. Rev. D",
	VOLUME=	"110",
	PAGES=	"030001",
	YEAR=	"2024",
}

@ARTICLE{Cunqueiro:2018jbh,
	AUTHOR=	"Cunqueiro, Leticia and P\l{}osko{\'n}, Mateusz",
	TITLE=	"Searching for the dead cone effects with iterative declustering of heavy-flavor jets",
	EPRINT=	"1812.00102",
	ARCHIVEPREFIX=	"arXiv",
	PRIMARYCLASS=	"hep-ph",
	DOI=	"10.1103/PhysRevD.99.074027",
	JOURNAL=	"Phys. Rev. D",
	VOLUME=	"99",
	PAGES=	"074027",
	YEAR=	"2019",
}

@ARTICLE{Ghira:2025nym,
	AUTHOR=	"Ghira, Andrea and Marzani, Simone and Soyez, Gregory",
	TITLE=	"{The Lund b jet plane}",
	EPRINT=	"2512.17408",
	ARCHIVEPREFIX=	"arXiv",
	PRIMARYCLASS=	"hep-ph",
	DOI=	"10.1007/JHEP05(2026)247",
	JOURNAL=	"JHEP",
	VOLUME=	"05",
	PAGES=	"247",
	YEAR=	"2026",
}

@ARTICLE{DELPHI:2000edu,
	AUTHOR=	"Abreu, P. and others",
	COLLABORATION=	"DELPHI",
	TITLE=	"Hadronization properties of b quarks compared to light quarks in e$^+$ e$^-$ $\to \mathrm{q} \overline{\mathrm{q}}$ from 183 to 200 {GeV}",
	EPRINT=	"hep-ex/0103022",
	ARCHIVEPREFIX=	"arXiv",
	DOI=	"10.1016/S0370-2693(00)00312-9",
	JOURNAL=	"Phys. Lett. B",
	VOLUME=	"479",
	PAGES=	"118",
	YEAR=	"2000",
}

@ARTICLE{CDF:2008ixu,
	AUTHOR=	"Aaltonen, T. and others",
	COLLABORATION=	"CDF",
	TITLE=	"{Measurement of b jet shapes in inclusive jet production in $\mathrm{p}\bar{\mathrm{p}}$ collisions at $\sqrt{s} = 1.96$\,{TeV}}",
	EPRINT=	"0806.1699",
	ARCHIVEPREFIX=	"arXiv",
	PRIMARYCLASS=	"hep-ex",
	REPORTNUMBER=	"FERMILAB-PUB-08-162-E",
	DOI=	"10.1103/PhysRevD.78.072005",
	JOURNAL=	"Phys. Rev. D",
	VOLUME=	"78",
	PAGES=	"072005",
	YEAR=	"2008",
}

@ARTICLE{ATLAS:2013uet,
	AUTHOR=	"{ATLAS Collaboration}",
	TITLE=	"Measurement of jet shapes in top quark pair events at $\sqrt{s}$ = 7 {TeV} using the {ATLAS} detector",
	EPRINT=	"1307.5749",
	ARCHIVEPREFIX=	"arXiv",
	PRIMARYCLASS=	"hep-ex",
	REPORTNUMBER=	"CERN-PH-EP-2013-067",
	DOI=	"10.1140/epjc/s10052-013-2676-3",
	JOURNAL=	"Eur. Phys. J. C",
	VOLUME=	"73",
	PAGES=	"2676",
	YEAR=	"2013",
}

@ARTICLE{Kluth:2023umf,
	AUTHOR=	"Kluth, Stefan and Ochs, Wolfgang and Perez Ramos, Redamy",
	TITLE=	"{Observation of the dead cone effect in charm and bottom quark jets and its QCD explanation}",
	EPRINT=	"2303.13343",
	ARCHIVEPREFIX=	"arXiv",
	PRIMARYCLASS=	"hep-ph",
	REPORTNUMBER=	"MPP-2023-52",
	DOI=	"10.1103/PhysRevD.107.094039",
	JOURNAL=	"Phys. Rev. D",
	VOLUME=	"107",
	PAGES=	"094039",
	YEAR=	"2023",
}

@ARTICLE{ALICE:2021aqk,
	AUTHOR=	"{ALICE Collaboration}",
	TITLE=	"Direct observation of the dead-cone effect in quantum chromodynamics",
	EPRINT=	"2106.05713",
	ARCHIVEPREFIX=	"arXiv",
	PRIMARYCLASS=	"nucl-ex",
	REPORTNUMBER=	"CERN-EP-2021-107",
	DOI=	"10.1038/s41586-022-04572-w",
	JOURNAL=	"Nature",
	VOLUME=	"605",
	PAGES=	"440",
	YEAR=	"2022",
	NOTE=	"[Erratum: \DOI{10.1038/s41586-022-05026-z}]",
}

@ARTICLE{LHCb:2025mcq,
	AUTHOR=	"{LHCb Collaboration}",
	TITLE=	"Measurement of the {Lund} plane for light- and beauty-quark jets",
	EPRINT=	"2505.23530",
	ARCHIVEPREFIX=	"arXiv",
	PRIMARYCLASS=	"hep-ex",
	REPORTNUMBER=	"LHCb-PAPER-2025-010, CERN-EP-2025-093",
	DOI=	"10.1103/r16d-b4my",
	JOURNAL=	"Phys. Rev. D",
	VOLUME=	"112",
	PAGES=	"072015",
	YEAR=	"2025",
}

@ARTICLE{CMS:2025dsz,
	AUTHOR=	"{CMS Collaboration}",
	TITLE=	"{Jet fragmentation function and groomed substructure of bottom quark jets in proton-proton collisions at 5.02 TeV}",
	EPRINT=	"2511.10666",
	ARCHIVEPREFIX=	"arXiv",
	PRIMARYCLASS=	"hep-ex",
	REPORTNUMBER=	"CMS-HIN-24-005, CERN-EP-2025-058",
	DOI=	"10.1007/JHEP04(2026)147",
	JOURNAL=	"JHEP",
	VOLUME=	"04",
	PAGES=	"147",
	YEAR=	"2026",
}

@ARTICLE{CMS:2025afq,
	AUTHOR=	"{CMS Collaboration}",
	TITLE=	"{Exploring small-angle emissions in charm quark jets in proton-proton collisions at $ \sqrt{s}=5.02 $ TeV}",
	EPRINT=	"2507.13469",
	ARCHIVEPREFIX=	"arXiv",
	PRIMARYCLASS=	"nucl-ex",
	REPORTNUMBER=	"CMS-HIN-24-007, CERN-EP-2025-089",
	DOI=	"10.1007/JHEP05(2026)176",
	JOURNAL=	"JHEP",
	VOLUME=	"05",
	PAGES=	"176",
	YEAR=	"2026",
}

@TECHREPORT{CMS-PAS-LUM-20-001,
	TITLE=	"Precision luminosity measurement in proton-proton collisions at 13 {TeV} with the {CMS} detector",
	AUTHOR=	"{CMS Collaboration}",
	URL=	"https://cds.cern.ch/record/2940794",
	TYPE=	"CMS Physics Analysis Summary",
	NUMBER=	"CMS-PAS-LUM-20-001",
	YEAR=	"2025",
}

@ARTICLE{Andersson:1988gp,
	AUTHOR=	"Andersson, Bo and Gustafson, Gosta and L{\"{o}}nnblad, Leif and Pettersson, Ulf",
	TITLE=	"Coherence Effects in Deep Inelastic Scattering",
	REPORTNUMBER=	"LU-TP-88-14",
	DOI=	"10.1007/BF01550942",
	JOURNAL=	"Z. Phys. C",
	VOLUME=	"43",
	PAGES=	"625",
	YEAR=	"1989",
}

@ARTICLE{Dreyer:2018nbf,
	AUTHOR=	"Dreyer, Fr{\'e}d{\'e}ric A. and Salam, Gavin P. and Soyez, Gr{\'e}gory",
	TITLE=	"The {Lund} Jet Plane",
	EPRINT=	"1807.04758",
	ARCHIVEPREFIX=	"arXiv",
	PRIMARYCLASS=	"hep-ph",
	REPORTNUMBER=	"CERN-TH-2018-151",
	DOI=	"10.1007/JHEP12(2018)064",
	JOURNAL=	"JHEP",
	VOLUME=	"12",
	PAGES=	"064",
	YEAR=	"2018",
}

@TECHREPORT{ATLAS:2025drf,
	TITLE=	"Explaining the {ATLAS} {GN2} flavor tagging algorithm with integrated gradients",
	AUTHOR=	"{ATLAS Collaboration}",
	NUMBER=	"ATL-PHYS-PUB-2025-029",
	YEAR=	"2025",
	TYPE=	"ATLAS Public Note",
	URL=	"https://cds.cern.ch/record/2936634",
}

@ARTICLE{CMS:2017wtu,
	AUTHOR=	"{CMS Collaboration}",
	TITLE=	"Identification of heavy-flavor jets with the {CMS} detector in pp collisions at 13 {TeV}",
	EPRINT=	"1712.07158",
	ARCHIVEPREFIX=	"arXiv",
	PRIMARYCLASS=	"physics.ins-det",
	REPORTNUMBER=	"CMS-BTV-16-002, CERN-EP-2017-326",
	DOI=	"10.1088/1748-0221/13/05/P05011",
	JOURNAL=	"JINST",
	VOLUME=	"13",
	PAGES=	"P05011",
	YEAR=	"2018",
}

@ARTICLE{Khachatryan:2010xn,
	AUTHOR=	"{CMS Collaboration}",
	TITLE=	"Measurements of Inclusive {W} and {Z} Cross Sections in {pp} Collisions at $\sqrt{s}=7$ {TeV}",
	EPRINT=	"1012.2466",
	ARCHIVEPREFIX=	"arXiv",
	PRIMARYCLASS=	"hep-ex",
	REPORTNUMBER=	"CERN-PH-EP-2010-050, CMS-EWK-10-002",
	DOI=	"10.1007/JHEP01(2011)080",
	JOURNAL=	"JHEP",
	VOLUME=	"01",
	PAGES=	"080",
	YEAR=	"2011",
}

@ARTICLE{CMS:2018rym,
	AUTHOR=	"{CMS Collaboration}",
	TITLE=	"Performance of the {CMS} muon detector and muon reconstruction with proton-proton collisions at $\sqrt{s}=$ 13 {TeV}",
	EPRINT=	"1804.04528",
	ARCHIVEPREFIX=	"arXiv",
	PRIMARYCLASS=	"physics.ins-det",
	REPORTNUMBER=	"CMS-MUO-16-001, CERN-EP-2018-058",
	DOI=	"10.1088/1748-0221/13/06/P06015",
	JOURNAL=	"JINST",
	VOLUME=	"13",
	PAGES=	"P06015",
	YEAR=	"2018",
}

@ARTICLE{CMS:2020uim,
	AUTHOR=	"{CMS Collaboration}",
	TITLE=	"Electron and photon reconstruction and identification with the {CMS} experiment at the {CERN} {LHC}",
	EPRINT=	"2012.06888",
	ARCHIVEPREFIX=	"arXiv",
	PRIMARYCLASS=	"hep-ex",
	REPORTNUMBER=	"CMS-EGM-17-001, CERN-EP-2020-219",
	DOI=	"10.1088/1748-0221/16/05/P05014",
	JOURNAL=	"JINST",
	VOLUME=	"16",
	PAGES=	"P05014",
	YEAR=	"2021",
}

@ARTICLE{ATLAS:2022miz,
	AUTHOR=	"{ATLAS Collaboration}",
	TITLE=	"{Measurements of jet observables sensitive to b quark fragmentation in $\mathrm{t\bar{t}}$ events at the LHC with the ATLAS detector}",
	EPRINT=	"2202.13901",
	ARCHIVEPREFIX=	"arXiv",
	PRIMARYCLASS=	"hep-ex",
	REPORTNUMBER=	"CERN-EP-2021-116",
	DOI=	"10.1103/PhysRevD.106.032008",
	JOURNAL=	"Phys. Rev. D",
	VOLUME=	"106",
	PAGES=	"032008",
	YEAR=	"2022",
}

@ARTICLE{ATLAS:2020bbn,
	AUTHOR=	"{ATLAS Collaboration}",
	TITLE=	"Measurement of the {Lund} Jet Plane Using Charged Particles in 13 {TeV} Proton-Proton Collisions with the {ATLAS} Detector",
	EPRINT=	"2004.03540",
	ARCHIVEPREFIX=	"arXiv",
	PRIMARYCLASS=	"hep-ex",
	REPORTNUMBER=	"CERN-EP-2020-030",
	DOI=	"10.1103/PhysRevLett.124.222002",
	JOURNAL=	"Phys. Rev. Lett.",
	VOLUME=	"124",
	PAGES=	"222002",
	YEAR=	"2020",
}

@ARTICLE{Havener:2021yhb,
	AUTHOR=	"{ALICE Collaboration}",
	TITLE=	"Measurement of the primary {Lund} jet plane density in pp collisions at $\sqrt{s} = \rm{13}$ {TeV} with {ALICE}",
	EPRINT=	"2111.00020",
	ARCHIVEPREFIX=	"arXiv",
	PRIMARYCLASS=	"hep-ex",
	DOI=	"10.22323/1.398.0364",
	JOURNAL=	"PoS",
	VOLUME=	"EPS-HEP2021",
	PAGES=	"364",
	YEAR=	"2022",
}

@ARTICLE{CMS:2023lpp,
	AUTHOR=	"{CMS Collaboration}",
	TITLE=	"Measurement of the primary {Lund} jet plane density in proton-proton collisions at $ \sqrt{\textrm{s}} $ = 13 {TeV}",
	EPRINT=	"2312.16343",
	ARCHIVEPREFIX=	"arXiv",
	PRIMARYCLASS=	"hep-ex",
	REPORTNUMBER=	"CMS-SMP-22-007, CERN-EP-2023-282",
	DOI=	"10.1007/JHEP05(2024)116",
	JOURNAL=	"JHEP",
	VOLUME=	"05",
	PAGES=	"116",
	YEAR=	"2024",
}

@ARTICLE{ATLAS:2024wrd,
	AUTHOR=	"{ATLAS Collaboration}",
	TITLE=	"Measurements of {Lund} subjet multiplicities in 13 {TeV} proton-proton collisions with the {ATLAS} detector",
	EPRINT=	"2402.13052",
	ARCHIVEPREFIX=	"arXiv",
	PRIMARYCLASS=	"hep-ex",
	REPORTNUMBER=	"CERN-EP-2024-029",
	DOI=	"10.1016/j.physletb.2024.139090",
	JOURNAL=	"Phys. Lett. B",
	VOLUME=	"859",
	PAGES=	"139090",
	YEAR=	"2024",
}

@ARTICLE{Cacciari:2008gp,
	AUTHOR=	"Cacciari, Matteo and Salam, Gavin P. and Soyez, Gregory",
	TITLE=	"The anti-$k_t$ jet clustering algorithm",
	JOURNAL=	"JHEP",
	VOLUME=	"04",
	YEAR=	"2008",
	PAGES=	"063",
	DOI=	"10.1088/1126-6708/2008/04/063",
	EPRINT=	"0802.1189",
	ARCHIVEPREFIX=	"arXiv",
	PRIMARYCLASS=	"hep-ph",
	REPORTNUMBER=	"LPTHE-07-03",
	SLACCITATION=	"%%CITATION = ARXIV:0802.1189;%%",
}

@ARTICLE{Dokshitzer:1997in,
	AUTHOR=	"Dokshitzer, Yuri L. and Leder, G. D. and Moretti, S. and Webber, B. R.",
	TITLE=	"Better jet clustering algorithms",
	EPRINT=	"hep-ph/9707323",
	ARCHIVEPREFIX=	"arXiv",
	REPORTNUMBER=	"CAVENDISH-HEP-97-06",
	DOI=	"10.1088/1126-6708/1997/08/001",
	JOURNAL=	"JHEP",
	VOLUME=	"08",
	PAGES=	"001",
	YEAR=	"1997",
}

@ARTICLE{Dasgupta:2020fwr,
	AUTHOR=	"Dasgupta, Mrinal and Dreyer, Fr{\'e}d{\'e}ric A. and Hamilton, Keith and Monni, Pier Francesco and Salam, Gavin P. and Soyez, Gregory",
	TITLE=	"Parton showers beyond leading logarithmic accuracy",
	EPRINT=	"2002.11114",
	ARCHIVEPREFIX=	"arXiv",
	PRIMARYCLASS=	"hep-ph",
	REPORTNUMBER=	"CERN-TH-2020-026",
	DOI=	"10.1103/PhysRevLett.125.052002",
	JOURNAL=	"Phys. Rev. Lett.",
	VOLUME=	"125",
	PAGES=	"052002",
	YEAR=	"2020",
}

@ARTICLE{Lifson:2020gua,
	AUTHOR=	"Lifson, Andrew and Salam, Gavin P. and Soyez, Gregory",
	TITLE=	"Calculating the primary {Lund} Jet Plane density",
	EPRINT=	"2007.06578",
	ARCHIVEPREFIX=	"arXiv",
	PRIMARYCLASS=	"hep-ph",
	DOI=	"10.1007/JHEP10(2020)170",
	JOURNAL=	"JHEP",
	VOLUME=	"10",
	PAGES=	"170",
	YEAR=	"2020",
}

@ARTICLE{Buckley:2010ar,
	AUTHOR=	"Buckley, Andy and Butterworth, Jonathan and Grellscheid, David and Hoeth, Hendrik and Lonnblad, Leif and Monk, James and Schulz, Holger and Siegert, Frank",
	TITLE=	"\textsc{rivet} user manual",
	EPRINT=	"1003.0694",
	ARCHIVEPREFIX=	"arXiv",
	PRIMARYCLASS=	"hep-ph",
	REPORTNUMBER=	"MCNET-10-03, MCnet/10/03",
	DOI=	"10.1016/j.cpc.2013.05.021",
	JOURNAL=	"Comput. Phys. Commun.",
	VOLUME=	"184",
	PAGES=	"2803",
	YEAR=	"2013",
}

@MISC{hepdata,
	TITLE=	"{HEPD}ata record for this analysis",
	DOI=	"10.17182/hepdata.180672",
	YEAR=	"2026",
}

@ARTICLE{CMS:2008xjf,
	AUTHOR=	"{CMS Collaboration}",
	TITLE=	"The {CMS} experiment at the {CERN} {LHC}",
	DOI=	"10.1088/1748-0221/3/08/S08004",
	JOURNAL=	"JINST",
	VOLUME=	"3",
	PAGES=	"S08004",
	YEAR=	"2008",
}

@ARTICLE{CMS:2023gfb,
	AUTHOR=	"{CMS Collaboration}",
	TITLE=	"Development of the {CMS} detector for the {CERN} {LHC} {Run 3}",
	EPRINT=	"2309.05466",
	ARCHIVEPREFIX=	"arXiv",
	PRIMARYCLASS=	"physics.ins-det",
	REPORTNUMBER=	"CMS-PRF-21-001, CERN-EP-2023-136",
	DOI=	"10.1088/1748-0221/19/05/P05064",
	JOURNAL=	"JINST",
	VOLUME=	"19",
	PAGES=	"P05064",
	YEAR=	"2024",
}

@ARTICLE{CMS:2020cmk,
	AUTHOR=	"{CMS Collaboration}",
	TITLE=	"Performance of the {CMS} {Level-1} trigger in proton-proton collisions at $\sqrt{s} = 13$\,{TeV}",
	JOURNAL=	"JINST",
	VOLUME=	"15",
	PAGES=	"P10017",
	YEAR=	"2020",
	DOI=	"10.1088/1748-0221/15/10/P10017",
	EPRINT=	"2006.10165",
	ARCHIVEPREFIX=	"arXiv",
	PRIMARYCLASS=	"hep-ex",
	REPORTNUMBER=	"CMS-TRG-17-001, CERN-EP-2020-065",
}

@ARTICLE{CMS:2016ngn,
	AUTHOR=	"{CMS Collaboration}",
	TITLE=	"The {CMS} trigger system",
	JOURNAL=	"JINST",
	VOLUME=	"12",
	PAGES=	"P01020",
	DOI=	"10.1088/1748-0221/12/01/P01020",
	YEAR=	"2017",
	EPRINT=	"1609.02366",
	ARCHIVEPREFIX=	"arXiv",
	PRIMARYCLASS=	"physics.ins-det",
	REPORTNUMBER=	"CMS-TRG-12-001, CERN-EP-2016-160",
	SLACCITATION=	"%%CITATION = ARXIV:1609.02366;%%",
}

@ARTICLE{CMS:2024aqx,
	AUTHOR=	"{CMS Collaboration}",
	TITLE=	"Performance of the {CMS} high-level trigger during {LHC} {Run 2}",
	EPRINT=	"2410.17038",
	ARCHIVEPREFIX=	"arXiv",
	PRIMARYCLASS=	"physics.ins-det",
	REPORTNUMBER=	"CMS-TRG-19-001, CERN-EP-2024-259",
	DOI=	"10.1088/1748-0221/19/11/P11021",
	JOURNAL=	"JINST",
	VOLUME=	"19",
	PAGES=	"P11021",
	YEAR=	"2024",
}

@ARTICLE{CMS:2014pgm,
	AUTHOR=	"{CMS Collaboration}",
	TITLE=	"Description and performance of track and primary-vertex reconstruction with the {CMS} tracker",
	EPRINT=	"1405.6569",
	ARCHIVEPREFIX=	"arXiv",
	PRIMARYCLASS=	"physics.ins-det",
	REPORTNUMBER=	"CMS-TRK-11-001, CERN-PH-EP-2014-070",
	DOI=	"10.1088/1748-0221/9/10/P10009",
	JOURNAL=	"JINST",
	VOLUME=	"9",
	PAGES=	"P10009",
	YEAR=	"2014",
}

@ARTICLE{CMS:2017yfk,
	AUTHOR=	"{CMS Collaboration}",
	TITLE=	"Particle-flow reconstruction and global event description with the {CMS} detector",
	EPRINT=	"1706.04965",
	ARCHIVEPREFIX=	"arXiv",
	PRIMARYCLASS=	"physics.ins-det",
	REPORTNUMBER=	"CMS-PRF-14-001, CERN-EP-2017-110",
	DOI=	"10.1088/1748-0221/12/10/P10003",
	JOURNAL=	"JINST",
	VOLUME=	"12",
	PAGES=	"P10003",
	YEAR=	"2017",
}

@TECHREPORT{CMS-TDR-15-02,
	AUTHOR=	"{CMS Collaboration}",
	TITLE=	"Technical proposal for the {Phase-II} upgrade of the {Compact Muon Solenoid}",
	URL=	"http://cds.cern.ch/record/2020886",
	TYPE=	"CMS Technical Proposal",
	NUMBER=	"CERN-LHCC-2015-010, CMS-TDR-15-02",
	YEAR=	"2015",
}

@ARTICLE{CMS:2020ebo,
	AUTHOR=	"{CMS Collaboration}",
	TITLE=	"Pileup mitigation at {CMS} in 13 {TeV} data",
	EPRINT=	"2003.00503",
	ARCHIVEPREFIX=	"arXiv",
	PRIMARYCLASS=	"hep-ex",
	REPORTNUMBER=	"CMS-JME-18-001, CERN-EP-2020-017",
	DOI=	"10.1088/1748-0221/15/09/P09018",
	JOURNAL=	"JINST",
	VOLUME=	"15",
	PAGES=	"P09018",
	YEAR=	"2020",
}

@ARTICLE{CMS:2011yuk,
	AUTHOR=	"{CMS Collaboration}",
	TITLE=	"{Measurement of $\mathrm{B}\bar{\mathrm{B}}$ angular correlations based on secondary vertex reconstruction at $\sqrt{s}=7$ TeV}",
	EPRINT=	"1102.3194",
	ARCHIVEPREFIX=	"arXiv",
	PRIMARYCLASS=	"hep-ex",
	REPORTNUMBER=	"CERN-PH-EP-10-093, CMS-BPH-10-010",
	DOI=	"10.1007/JHEP03(2011)136",
	JOURNAL=	"JHEP",
	VOLUME=	"03",
	PAGES=	"136",
	YEAR=	"2011",
}

@ARTICLE{Cacciari:2011ma,
	AUTHOR=	"Cacciari, Matteo and Salam, Gavin P. and Soyez, Gregory",
	TITLE=	"FastJet User Manual",
	EPRINT=	"1111.6097",
	ARCHIVEPREFIX=	"arXiv",
	PRIMARYCLASS=	"hep-ph",
	REPORTNUMBER=	"CERN-PH-TH-2011-297",
	DOI=	"10.1140/epjc/s10052-012-1896-2",
	JOURNAL=	"Eur. Phys. J. C",
	VOLUME=	"72",
	PAGES=	"1896",
	YEAR=	"2012",
}

@ARTICLE{CMS:2016lmd,
	AUTHOR=	"{CMS Collaboration}",
	TITLE=	"Jet energy scale and resolution in the {CMS} experiment in pp collisions at 8 {TeV}",
	EPRINT=	"1607.03663",
	ARCHIVEPREFIX=	"arXiv",
	PRIMARYCLASS=	"hep-ex",
	REPORTNUMBER=	"CMS-JME-13-004, CERN-PH-EP-2015-305",
	DOI=	"10.1088/1748-0221/12/02/P02014",
	JOURNAL=	"JINST",
	VOLUME=	"12",
	PAGES=	"P02014",
	YEAR=	"2017",
}

@ARTICLE{Cacciari:2007fd,
	AUTHOR=	"Cacciari, Matteo and Salam, Gavin P.",
	TITLE=	"Pileup subtraction using jet areas",
	EPRINT=	"0707.1378",
	ARCHIVEPREFIX=	"arXiv",
	PRIMARYCLASS=	"hep-ph",
	REPORTNUMBER=	"LPTHE-07-01",
	DOI=	"10.1016/j.physletb.2007.09.077",
	JOURNAL=	"Phys. Lett. B",
	VOLUME=	"659",
	PAGES=	"119",
	YEAR=	"2008",
}

@ARTICLE{ATLAS:2024jrp,
	AUTHOR=	"{ATLAS Collaboration}",
	TITLE=	"Measurement of jet track functions in pp collisions at $\sqrt{s}=$ 13 {TeV} with the {ATLAS} detector",
	EPRINT=	"2502.02062",
	ARCHIVEPREFIX=	"arXiv",
	PRIMARYCLASS=	"hep-ex",
	REPORTNUMBER=	"CERN-EP-2024-333",
	DOI=	"10.1016/j.physletb.2025.139680",
	JOURNAL=	"Phys. Lett. B",
	VOLUME=	"868",
	PAGES=	"139680",
	YEAR=	"2025",
}

@ARTICLE{Nason:2004rx,
	AUTHOR=	"Nason, Paolo",
	TITLE=	"A New method for combining {NLO QCD} with shower {M}onte {C}arlo algorithms",
	EPRINT=	"hep-ph/0409146",
	ARCHIVEPREFIX=	"arXiv",
	REPORTNUMBER=	"BICOCCA-FT-04-11",
	DOI=	"10.1088/1126-6708/2004/11/040",
	JOURNAL=	"JHEP",
	VOLUME=	"11",
	PAGES=	"040",
	YEAR=	"2004",
}

@ARTICLE{Frixione:2007vw,
	AUTHOR=	"Frixione, Stefano and Nason, Paolo and Oleari, Carlo",
	TITLE=	"Matching {NLO QCD} computations with Parton Shower simulations: the \textsc{powheg} method",
	EPRINT=	"0709.2092",
	ARCHIVEPREFIX=	"arXiv",
	PRIMARYCLASS=	"hep-ph",
	REPORTNUMBER=	"BICOCCA-FT-07-9, GEF-TH-21-2007",
	DOI=	"10.1088/1126-6708/2007/11/070",
	JOURNAL=	"JHEP",
	VOLUME=	"11",
	PAGES=	"070",
	YEAR=	"2007",
}

@ARTICLE{Alioli:2010xd,
	AUTHOR=	"Alioli, Simone and Nason, Paolo and Oleari, Carlo and Re, Emanuele",
	TITLE=	"A general framework for implementing {NLO} calculations in shower {M}onte {C}arlo programs: the \textsc{powheg box}",
	EPRINT=	"1002.2581",
	ARCHIVEPREFIX=	"arXiv",
	PRIMARYCLASS=	"hep-ph",
	REPORTNUMBER=	"DESY-10-018, SFB-CPP-10-22, IPPP-10-11, DCPT-10-22",
	DOI=	"10.1007/JHEP06(2010)043",
	JOURNAL=	"JHEP",
	VOLUME=	"06",
	PAGES=	"043",
	YEAR=	"2010",
}

@ARTICLE{Alioli:2010xa,
	AUTHOR=	"Alioli, Simone and Hamilton, Keith and Nason, Paolo and Oleari, Carlo and Re, Emanuele",
	TITLE=	"Jet pair production in \textsc{powheg}",
	EPRINT=	"1012.3380",
	ARCHIVEPREFIX=	"arXiv",
	PRIMARYCLASS=	"hep-ph",
	REPORTNUMBER=	"DESY-10-233, SFB-CPP-10-128, IPPP-10-100, DCPT-10-200, MCNET-10-23",
	DOI=	"10.1007/JHEP04(2011)081",
	JOURNAL=	"JHEP",
	VOLUME=	"04",
	PAGES=	"081",
	YEAR=	"2011",
}

@ARTICLE{Alioli:2008tz,
	AUTHOR=	"Alioli, Simone and Nason, Paolo and Oleari, Carlo and Re, Emanuele",
	TITLE=	"{NLO} {Higgs} boson production via gluon fusion matched with shower in \textsc{powheg}",
	EPRINT=	"0812.0578",
	ARCHIVEPREFIX=	"arXiv",
	PRIMARYCLASS=	"hep-ph",
	DOI=	"10.1088/1126-6708/2009/04/002",
	JOURNAL=	"JHEP",
	VOLUME=	"04",
	PAGES=	"002",
	YEAR=	"2009",
}

@ARTICLE{Bagnaschi:2011tu,
	AUTHOR=	"Bagnaschi, E. and Degrassi, G. and Slavich, P. and Vicini, A.",
	TITLE=	"{Higgs} production via gluon fusion in the \textsc{powheg} approach in the {SM} and in the {MSSM}",
	EPRINT=	"1111.2854",
	ARCHIVEPREFIX=	"arXiv",
	PRIMARYCLASS=	"hep-ph",
	REPORTNUMBER=	"RM3-TH-11-5, IFUM-990-FT",
	DOI=	"10.1007/JHEP02(2012)088",
	JOURNAL=	"JHEP",
	VOLUME=	"02",
	PAGES=	"088",
	YEAR=	"2012",
}

@ARTICLE{Heinrich:2017kxx,
	AUTHOR=	"Heinrich, Gudrun and Jones, Stephen and Kerner, Matthias and Luisoni, Gionata and Vryonidou, Eleni",
	TITLE=	"{NLO} predictions for {Higgs} boson pair production with full top quark mass dependence matched to parton showers",
	EPRINT=	"1703.09252",
	ARCHIVEPREFIX=	"arXiv",
	PRIMARYCLASS=	"hep-ph",
	REPORTNUMBER=	"CERN-TH-2017-069, MPP-2017-41, NIKHEF-2017-020",
	DOI=	"10.1007/JHEP08(2017)088",
	JOURNAL=	"JHEP",
	VOLUME=	"08",
	PAGES=	"088",
	YEAR=	"2017",
}

@ARTICLE{Buchalla:2018yce,
	AUTHOR=	"Buchalla, G. and Capozi, M. and Celis, A. and Heinrich, G. and Scyboz, L.",
	TITLE=	"{Higgs} boson pair production in non-linear Effective Field Theory with full $m_\mathrm{t}$-dependence at {NLO QCD}",
	EPRINT=	"1806.05162",
	ARCHIVEPREFIX=	"arXiv",
	PRIMARYCLASS=	"hep-ph",
	REPORTNUMBER=	"LMU-ASC 34/18, MPP-2018-127, LMU-ASC-34-18",
	DOI=	"10.1007/JHEP09(2018)057",
	JOURNAL=	"JHEP",
	VOLUME=	"09",
	PAGES=	"057",
	YEAR=	"2018",
}

@ARTICLE{Frederix:2012ps,
	AUTHOR=	"Frederix, Rikkert and Frixione, Stefano",
	TITLE=	"Merging meets matching in \textsc{mc@nlo}",
	EPRINT=	"1209.6215",
	ARCHIVEPREFIX=	"arXiv",
	PRIMARYCLASS=	"hep-ph",
	REPORTNUMBER=	"CERN-PH-TH-2012-247, ZU-TH-21-12",
	DOI=	"10.1007/JHEP12(2012)061",
	JOURNAL=	"JHEP",
	VOLUME=	"12",
	PAGES=	"061",
	YEAR=	"2012",
}

@ARTICLE{Sjostrand:2014zea,
	AUTHOR=	"Sj{\"{o}}strand, Torbj{\"{o}}rn and Ask, Stefan and Christiansen, Jesper R. and Corke, Richard and Desai, Nishita and Ilten, Philip and Mrenna, Stephen and Prestel, Stefan and Rasmussen, Christine O. and Skands, Peter Z.",
	TITLE=	"An introduction to \textsc{pythia} 8.2",
	EPRINT=	"1410.3012",
	ARCHIVEPREFIX=	"arXiv",
	PRIMARYCLASS=	"hep-ph",
	REPORTNUMBER=	"LU-TP-14-36, MCNET-14-22, CERN-PH-TH-2014-190, FERMILAB-PUB-14-316-CD, DESY-14-178, SLAC-PUB-16122",
	DOI=	"10.1016/j.cpc.2015.01.024",
	JOURNAL=	"Comput. Phys. Commun.",
	VOLUME=	"191",
	PAGES=	"159",
	YEAR=	"2015",
}

@ARTICLE{Sirunyan:2019dfx,
	AUTHOR=	"{CMS Collaboration}",
	TITLE=	"Extraction and validation of a new set of {CMS} \textsc{pythia} 8 tunes from underlying-event measurements",
	EPRINT=	"1903.12179",
	ARCHIVEPREFIX=	"arXiv",
	PRIMARYCLASS=	"hep-ex",
	REPORTNUMBER=	"CMS-GEN-17-001, CERN-EP-2019-007",
	DOI=	"10.1140/epjc/s10052-019-7499-4",
	JOURNAL=	"Eur. Phys. J. C",
	VOLUME=	"80",
	PAGES=	"4",
	YEAR=	"2020",
}

@ARTICLE{Ball:2017nwa,
	AUTHOR=	"Ball, Richard D. and others",
	COLLABORATION=	"NNPDF {C}ollaboration",
	TITLE=	"Parton distributions from high-precision collider data",
	EPRINT=	"1706.00428",
	ARCHIVEPREFIX=	"arXiv",
	PRIMARYCLASS=	"hep-ph",
	REPORTNUMBER=	"CAVENDISH-HEP-17-06, CERN-TH-2017-077, EDINBURGH-2017-08, NIKHEF-2017-006, OUTP-17-04P, TIF-UNIMI-2017-3",
	DOI=	"10.1140/epjc/s10052-017-5199-5",
	JOURNAL=	"Eur. Phys. J. C",
	VOLUME=	"77",
	PAGES=	"663",
	YEAR=	"2017",
}

@ARTICLE{Bellm:2015jjp,
	AUTHOR=	"Bellm, Johannes and others",
	TITLE=	"{\textsc{herwig} 7.0/\textsc{herwig}++ 3.0} release note",
	JOURNAL=	"Eur. Phys. J. C",
	VOLUME=	"76",
	YEAR=	"2016",
	PAGES=	"196",
	DOI=	"10.1140/epjc/s10052-016-4018-8",
	EPRINT=	"1512.01178",
	ARCHIVEPREFIX=	"arXiv",
	PRIMARYCLASS=	"hep-ph",
	REPORTNUMBER=	"CERN-PH-TH-2015-289, MAN-HEP-2015-15, IFJPAN-IV-2015-13, KA-TP-18-2015, DCPT-15-142, MCNET-15-28, IPPP-15-71, HERWIG-2015-01",
	SLACCITATION=	"%%CITATION = ARXIV:1512.01178;%%",
}

@ARTICLE{CMS:2020dqt,
	AUTHOR=	"{CMS Collaboration}",
	TITLE=	"{D}evelopment and validation of \textsc{herwig} 7 tunes from {CMS} underlying-event measurements",
	EPRINT=	"2011.03422",
	ARCHIVEPREFIX=	"arXiv",
	PRIMARYCLASS=	"hep-ex",
	REPORTNUMBER=	"CMS-GEN-19-001, CERN-EP-2020-182",
	DOI=	"10.1140/epjc/s10052-021-08949-5",
	JOURNAL=	"Eur. Phys. J. C",
	VOLUME=	"81",
	PAGES=	"312",
	YEAR=	"2021",
}

@UNPUBLISHED{Bellm:2025pcw,
	AUTHOR=	"Bellm, J. and others",
	TITLE=	"{The physics of \textsc{herwig} 7}",
	EPRINT=	"2512.16645",
	ARCHIVEPREFIX=	"arXiv",
	PRIMARYCLASS=	"hep-ph",
	REPORTNUMBER=	"CERN-TH-2025-252, IPPP-25-57, HERWIG-2025-01, KA-TP-36-2025, MCNET-25-31",
	YEAR=	"2025",
}

@ARTICLE{Agostinelli:2002hh,
	AUTHOR=	"Agostinelli, S. and others",
	COLLABORATION=	"GEANT4",
	TITLE=	"{\GEANTfour}---a simulation toolkit",
	JOURNAL=	"Nucl. Instrum. Meth. A",
	VOLUME=	"506",
	YEAR=	"2003",
	PAGES=	"250",
	DOI=	"10.1016/S0168-9002(03)01368-8",
	SLACCITATION=	"%%CITATION = NUIMA,A506,250;%%",
}

@ARTICLE{Bols:2020bkb,
	AUTHOR=	"Bols, Emil and Kieseler, Jan and Verzetti, Mauro and Stoye, Markus and Stakia, Anna",
	TITLE=	"Jet flavor classification using \textsc{DeepJet}",
	EPRINT=	"2008.10519",
	ARCHIVEPREFIX=	"arXiv",
	PRIMARYCLASS=	"hep-ex",
	DOI=	"10.1088/1748-0221/15/12/P12012",
	JOURNAL=	"JINST",
	VOLUME=	"15",
	PAGES=	"P12012",
	YEAR=	"2020",
}

@INPROCEEDINGS{Vaswani:2017lxt,
	AUTHOR=	"Vaswani, Ashish and Shazeer, Noam and Parmar, Niki and Uszkoreit, Jakob and Jones, Llion and Gomez, Aidan N and Kaiser, \L ukasz and Polosukhin, Illia",
	BOOKTITLE=	"Advances in neural information processing systems",
	EDITOR=	"I. Guyon and U. Von Luxburg and S. Bengio and H. Wallach and R. Fergus and S. Vishwanathan and R. Garnett",
	PUBLISHER=	"Curran Associates, Inc.",
	TITLE=	"Attention is all you need",
	URL=	"https://proceedings.neurips.cc/paper_files/paper/2017/file/3f5ee243547dee91fbd053c1c4a845aa-Paper.pdf",
	VOLUME=	"30",
	YEAR=	"2017",
	ARCHIVEPREFIX=	"arXiv",
	EPRINT=	"1706.03762",
}

@ARTICLE{DAgostini:1994fjx,
	AUTHOR=	"D'Agostini, G.",
	TITLE=	"A multidimensional unfolding method based on {Bayes}' theorem",
	REPORTNUMBER=	"DESY-94-099",
	DOI=	"10.1016/0168-9002(95)00274-X",
	JOURNAL=	"Nucl. Instrum. Meth. A",
	VOLUME=	"362",
	PAGES=	"487",
	YEAR=	"1995",
}

@ARTICLE{Brenner:2019lmf,
	AUTHOR=	"Brenner, Lydia and Balasubramanian, Rahul and Burgard, Carsten and Verkerke, Wouter and Cowan, Glen and Verschuuren, Pim and Croft, Vincent",
	TITLE=	"Comparison of unfolding methods using \textsc{RooFitUnfold}",
	EPRINT=	"1910.14654",
	ARCHIVEPREFIX=	"arXiv",
	PRIMARYCLASS=	"physics.data-an",
	DOI=	"10.1142/S0217751X20501456",
	JOURNAL=	"Int. J. Mod. Phys. A",
	VOLUME=	"35",
	PAGES=	"2050145",
	YEAR=	"2020",
}

@INPROCEEDINGS{pvalue,
	AUTHOR=	"Demortier, Luc",
	TITLE=	"{P} values and nuisance parameters",
	BOOKTITLE=	"{\it Proc. Workshop on statistical issues for LHC physics, PHYSTAT-LHC\/}",
	YEAR=	"2007",
	PAGES=	"23",
	DOI=	"10.5170/CERN-2008-001",
	REPORTNUMBER=	"CERN-2008-001",
	SLACCITATION=	"%%CITATION = CERN-2008-001;%%",
}

@TECHREPORT{CMS-DP-2022-012,
	TITLE=	"{Tracking performances for charged pions with Run2 Legacy data}",
	AUTHOR=	"{CMS Collaboration}",
	TYPE=	"CMS Detector Performance Summary",
	YEAR=	"2022",
	URL=	"https://cds.cern.ch/record/2810814",
	NUMBER=	"CMS-DP-2022-012",
}

@ARTICLE{Butterworth:2015oua,
	AUTHOR=	"Butterworth, Jon and others",
	ARCHIVEPREFIX=	"arXiv",
	DOI=	"10.1088/0954-3899/43/2/023001",
	EPRINT=	"1510.03865",
	JOURNAL=	"J. Phys. G",
	PAGES=	"023001",
	PRIMARYCLASS=	"hep-ph",
	REPORTNUMBER=	"OUTP-15-17P, SMU-HEP-15-12, TIF-UNIMI-2015-14, LCTS-2015-27, CERN-PH-TH-2015-249",
	TITLE=	"{PDF4LHC} recommendations for {LHC} {Run II}",
	VOLUME=	"43",
	YEAR=	"2016",
}

@ARTICLE{PDF4LHCWorkingGroup:2022cjn,
	AUTHOR=	"Ball, Richard D. and others",
	COLLABORATION=	"PDF4LHC Working Group",
	TITLE=	"The {PDF4LHC21} combination of global {PDF} fits for the {LHC} {Run III}",
	EPRINT=	"2203.05506",
	ARCHIVEPREFIX=	"arXiv",
	PRIMARYCLASS=	"hep-ph",
	REPORTNUMBER=	"Edinburgh 2021/31, FERMILAB-PUB-22-121-QIS-SCD-T, MSUHEP-22-010, Nikhef 2021-033, SMU-HEP-22-01",
	DOI=	"10.1088/1361-6471/ac7216",
	JOURNAL=	"J. Phys. G",
	VOLUME=	"49",
	PAGES=	"080501",
	YEAR=	"2022",
}

@ARTICLE{Efron:1979bxm,
	AUTHOR=	"Efron, B.",
	TITLE=	"{Bootstrap methods: another look at the jackknife}",
	DOI=	"10.1214/aos/1176344552",
	JOURNAL=	"Annals Statist.",
	VOLUME=	"7",
	PAGES=	"1",
	YEAR=	"1979",
}

\appendix

\section{Details of the transformer encoder}
\label{app:endMat}

Details of the transformer encoder (TFE) used to identify charged \PQb hadron decay products in jets are discussed here.
Samples of \PQb jets with $40<\pt<200\GeV$ from the {\POWHEG v2.0 + \PYTHIA v8.240} dileptonic \ttbar samples are used for the training, testing, and validation of the TFE by splitting the samples into 80\%, 10\%, and 10\% fractions, respectively.
Each charged jet constituent is matched to a generator-level charged particle that is within $\sqrt{\smash[b]{(\Delta \eta)^2+(\Delta \phi)^2}}<0.02$ and with \pt within 20\% of difference.
According to the matched partner, the charged jet constituents are labeled as either signals from \PQb hadron decays or backgrounds.
About 5.6\% of the constituents are not matched to a generator-level particle. 
These unmatched particles are primarily from pileup vertices and are labeled as background.

\begin{figure}[htbp!]
  \centering
  \ifthenelse{\boolean{cms@external}}{
    \includegraphics[width=0.3\textwidth]{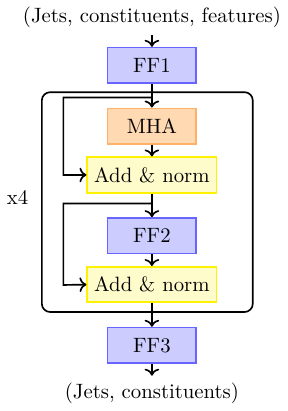}
  }{
    \includegraphics[width=0.4\textwidth]{Figure_004.pdf}
  }
  \caption{Architecture of the transformer encoder trained to identify \PQb hadron decay products in jets.
        Input to the network are three-dimensional arrays that represent the number of jets, the number of constituents in a jet, and the number of features.
	The first feed forward (FF1) block is a single-layered dense network that expands the feature dimension to 128. 
	This is followed by the multihead attention (MHA) block with four heads, and another FF block (FF2) with two layers. 
	Both blocks keep 128 hidden dimensions.
	Skip connections and layer normalizations (Add \& norm) are added in between the MHA and FF1, and after FF2. 
	The MHA, FF2, and Add \& norm blocks are repeated together for four different times. 
	The final block is a single-layered FF (FF3) that contracts the final output to two-dimensional arrays representing a sequence of probabilities for each constituent in a jet to be a \PQb hadron decay product or not.}
  \label{fig:tfe_arch} 
\end{figure}

The architecture of the TFE is illustrated in Fig.~\ref{fig:tfe_arch}, which follows closely the encoder structure described in Ref.~\cite{Vaswani:2017lxt}. 
Inputs to the TFE are three-dimensional arrays representing the number of jets, the number of constituents in a jet, and the number of input features. 
The input features to the TFE include the jet \pt and $\eta$, the constituents' $\eta$, their $\sqrt{\smash[b]{(\Delta\eta)^{2}+(\Delta\phi)^{2}}}$ separation from the jet axis, and their rapidities relative to the jet axis.
Each charged constituent is uniquely associated with a reconstructed track, and the track can in turn be associated with a SV. 
Details of the track and SV reconstruction are described in Ref.~\cite{CMS:2014pgm} and~\cite{CMS:2011yuk}, respectively.
Observables based on the associated tracks of the constituents are also added, including the transverse and longitudinal momentum fractions of the tracks relative to the jet axis, minimum approach distance to the jet axis, the distance from the PV to the point of closest approach between the track and the jet axis, the impact and transverse impact parameters of the tracks, and the same impact parameters divided by their uncertainties. 
Observables based on the associated SV of the tracks, including their transverse momentum and energy relative to the jet and the jet constituent, the $\chi^2$ in the SV fit per degree of freedom, the track multiplicity associated with an SV, the impact and transverse impact parameters divided by their uncertainties, and the corrected SV mass~\cite{SLD:1997ihw} are also used.
The corrected SV mass is defined as $\sqrt{\smash[b]{m_{\mathrm{SV}}^{2}+p_{\mathrm{SV}}^{2}\sin^{2}\theta}}+p_{\mathrm{SV}}\sin\theta$, where $m_{\mathrm{SV}}$ ($p_{\mathrm{SV}}$) is the invariant mass (momentum) of the SV, and $\theta$ is the angle between the direction of $p_{\mathrm{SV}}$ and the direction from the PV to the SV.
These features are normalized to have a mean of 0 and a standard deviation of 1. 
For constituents without an associated SV, the SV variables are set to be 10 standard deviations below the mean, far from any physical values. 
The network outputs a sequence of probabilities for each constituent in a jet to be from \PQb hadron decays or not.

The {\textsc{PyTorch} v2.4.0} software~\cite{Paszke:2019xhz,10.1145/3620665.3640366} is used in the training. 
The binary cross entropy loss is used with the \textsc{AdamW} gradient descent optimizer~\cite{Loshchilov:2017bsp}.
Parameters for the \textsc{AdamW} optimizer are left as the default values in \textsc{PyTorch}.
The validation loss is monitored to reduce the learning rate by half when it stops improving for 5 continuous epochs, and to terminate the training when it stops improving for 20 continuous epochs. 
A total of 100 epochs were trained before both the training and validation losses plateaued and stabilized.
The performance of the model, in terms of the area under the receiver operating characteristic curve (AUC), and the signal and background efficiencies are listed in Table~\ref{tab:tfeperf}. 
The signal and background efficiencies are defined by the fraction of correctly identified signal and background particles, respectively.
No significant difference in performance is observed between the training, testing, and validation sets. 

\begin{table*}[htbp]
\centering
\topcaption{Performance of the TFE to identify \PQb hadron decay products in jets in the training, testing, and validation sets of the \POWHEG{}+\PYTHIA dileptonic \ttbar sample.}
\label{tab:tfeperf}
\begin{scotch}{lccc}
Set                & AUC (\%) &  Signal efficiency (\%) &  Background efficiency (\%)   \\
\hline
Training           & 98.4     &  96.0                         &    91.7       \\
Testing            & 98.2     &  95.7                         &    91.2       \\
Validation         & 98.2     &  95.7                         &    91.3       \\
\end{scotch}
\end{table*}

\cleardoublepage \section{The CMS Collaboration \label{app:collab}}\begin{sloppypar}\hyphenpenalty=5000\widowpenalty=500\clubpenalty=5000\cmsinstitute{Yerevan Physics Institute, Yerevan, Armenia}
{\tolerance=6000
A.~Gevorgyan\cmsorcid{0000-0003-2751-9489}, A.~Hayrapetyan, V.~Makarenko\cmsorcid{0000-0002-8406-8605}, A.~Tumasyan\cmsAuthorMark{1}\cmsorcid{0009-0000-0684-6742}
\par}
\cmsinstitute{Marietta Blau Institute for Particle Physics, Vienna, Austria}
{\tolerance=6000
W.~Adam\cmsorcid{0000-0001-9099-4341}, L.~Benato\cmsorcid{0000-0001-5135-7489}, T.~Bergauer\cmsorcid{0000-0002-5786-0293}, M.~Dragicevic\cmsorcid{0000-0003-1967-6783}, S.~Gundacker\cmsorcid{0000-0003-2087-3266}, A.K.~Guven\cmsorcid{0009-0004-5670-5138}, P.S.~Hussain\cmsorcid{0000-0002-4825-5278}, M.~Jeitler\cmsAuthorMark{2}\cmsorcid{0000-0002-5141-9560}, N.~Krammer\cmsorcid{0000-0002-0548-0985}, A.~Li\cmsorcid{0000-0002-4547-116X}, D.~Liko\cmsorcid{0000-0002-3380-473X}, M.~Matthewman, J.~Schieck\cmsAuthorMark{2}\cmsorcid{0000-0002-1058-8093}, R.~Sch\"{o}fbeck\cmsAuthorMark{2}\cmsorcid{0000-0002-2332-8784}, M.~Shooshtari\cmsorcid{0009-0004-8882-4887}, M.~Sonawane\cmsorcid{0000-0003-0510-7010}, N.~Van~Den~Bossche\cmsorcid{0000-0003-2973-4991}, W.~Waltenberger\cmsorcid{0000-0002-6215-7228}, C.E.~Wulz\cmsAuthorMark{2}\cmsorcid{0000-0001-9226-5812}
% ...
\par}
$^{1}$Also at Yerevan State University, Yerevan, Armenia\\
$^{2}$Also at Technische Universit\"{a}t Wien, Vienna, Austria\\
% ...
\end{sloppypar}
\end{document}